\documentclass[12pt,preprint]{aastex}

\usepackage{graphics}

\shorttitle{SEGUE Parameter Pipeline}

\shortauthors{Lee et al.}

\begin{document}

\title{The SEGUE Stellar Parameter Pipeline. I. Description and Initial Validation Tests}

\author{Young Sun Lee, Timothy C. Beers, Thirupathi Sivarani}
\affil{Department of Physics $\&$ Astronomy,
\\CSCE: Center for the Study of Cosmic Evolution, \\ and JINA: Joint Institute for Nuclear
Astrophysics, \\ Michigan State University, East Lansing, MI 48824,
USA} \email{lee@pa.msu.edu, beers@pa.msu.edu, thirupathi@pa.msu.edu}

\author{Carlos Allende Prieto, Lars Koesterke}
\affil{Department of Astronomy, \\ University of Texas, Austin, TX
78712} \email{callende@astro.as.utexas.edu}

\author{Ronald Wilhelm}
\affil{Department of Physics,\\ Texas Tech University, Lubbock, TX
79409} \email{ron.wilhelm@ttu.edu}

\author{John E. Norris}
\affil{Research School of Astronomy and Astrophysics, \\ The Australian
National University, Weston, ACT 2611, Australia}
\email{jen@mso.anu.edu.au}

\author{Coryn A. L. Bailer-Jones, Paola Re Fiorentin}
\affil{Max-Planck-Institute f\"ur Astronomy, K\"onigstuhl 17, D-69117, Heidelberg, Germany}
\email{calj@mpia-hd.mpg.de, fiorent@mpia-hd.mpg.de}

\author{Constance M. Rockosi}
\affil{Department of Astronomy,\\
University of California, Santa Cruz, CA 95064}
\email{crockosi@ucolick.org}

\author{Brian Yanny}
\affil{Fermi National Accelerator Laboratory,\\
Batavia, IL 60510} \email{yanny@fnal.gov}

\author{Heidi J. Newberg}
\affil{Department of Physics $\&$ Astronomy,\\
Rensselaer Polytechnical Institute, Troy, NY 12180}
\email{newbeh@rpi.edu}

\author{Kevin R. Covey}
\affil{Harvard-Smithsonian Center for Astrophysics,\\
60 Garden Street, Cambridge, MA 02138}
\email{kcovey@cfa.harvard.edu}

\begin{abstract}

We describe the development and implementation of the SEGUE (Sloan Extension for
Galactic Exploration and Understanding) Stellar Parameter Pipeline (SSPP). The
SSPP derives, using multiple techniques, radial velocities and the fundamental
stellar atmospheric parameters (effective temperature, surface gravity, and
metallicity) for AFGK-type stars, based on medium-resolution spectroscopy and
$ugriz$ photometry obtained during the course of the original Sloan Digital Sky
Survey (SDSS-I) and its Galactic extension (SDSS-II/SEGUE). The SSPP also
provides spectral classification for a much wider range of stars, including
stars with temperatures outside of the window where atmospheric parameters can
be estimated with the current approaches. This is Paper I in a series of papers
on the SSPP; it provides an overview of the SSPP, and initial tests of its
performance using multiple data sets. Random and systematic errors are
critically examined for the current version of the SSPP, which has been used for
the sixth public data release of the SDSS (DR-6).

\end{abstract}


\section{Introduction}

The Sloan Extension for Galactic Understanding and Exploration (SEGUE) is one of
three surveys that are being executed as part of the current extension of the
Sloan Digital Sky Survey (SDSS-II), which consists of LEGACY, SUPERNOVA SURVEY,
and SEGUE. The SEGUE program is designed to obtain $ugriz$ imaging of some 3500
square degrees of sky outside of the original SDSS-I footprint (Fukugita et al.
1996; Gunn et al. 1998, 2006; York et al. 2000; Stoughton et al. 2002; Abazajian
et al. 2003, 2004, 2005; Pier et al. 2003; Adelman-McCarthy et al. 2006, 2007a).
The regions of sky targeted are primarily at lower Galactic latitudes ($|b|~<
~35^{\circ}$), in order to better sample the disk/halo interface of the Milky
Way. As of Data Release 6 (DR-6, Adelman-McCarthy et al. 2007b), about 85\% of
the planned additional imaging has already been completed. SEGUE is also
obtaining $R$ $\simeq$ 2000 spectroscopy, over the wavelength range
$3800-9200$\, {\AA}, for some 250,000 stars in 200 selected areas over the sky
available from Apache Point, New Mexico. The spectroscopic candidates are
selected on the basis of $ugriz$ photometry to populate roughly 16 target
categories, chosen to explore the nature of the Galactic stellar populations at
distances from 0.5 kpc to over 100 kpc from the Sun. Spectroscopic observations
have been obtained for roughly half of the planned targets thus far, a total of
about 120,000 spectra. The SEGUE data clearly require automated analysis tools
in order to efficiently extract the maximum amount of useful astrophysical
information for the targeted stars, in particular their stellar atmospheric
parameters, over wide ranges of effective temperature ($T_{\rm eff}$),
surface gravity (log $g$), and metallicity ([Fe/H]).

Numerous methods have been developed in the past in order to extract
atmospheric-parameter estimates from medium-resolution stellar spectra in a
fast, efficient, and automated fashion. These approaches include techniques for
finding the minimum distance (parameterized in various ways) between observed
spectra and grids of synthetic spectra (e.g., Allende Prieto et al. 2006),
non-linear regression models (e.g., Re Fiorentin et al. 2007, and references
therein), correlations between broadband colors and the strength of prominent
metallic lines, such as the Ca~II K line (Beers et al. 1999), auto-correlation
analysis of a stellar spectrum (Beers et al. 1999, and references therein),
obtaining fits of spectral lines (or summed line indices) as a function of
broadband colors (Wilhelm et al. 1999), or the behavior of the Ca~II triplet
lines as a function of broadband color (Cenarro et al. 2001a,b). However, each
of these approaches exhibits optimal behavior only over restricted temperature
and metallicity ranges; outside of these regions they are often un-calibrated,
suffer from saturation of the metallic lines used in their estimates at high
metallicity or low temperatures, or lose efficacy due to the weakening of
metallic species at low metallicity or high temperatures. The methods that make
use of specific spectral features are susceptible to other problems, e.g., the
presence of emission in the core of the Ca~II K line for chromospherically
active stars, or poor telluric line subtraction in the region of the Ca~II
triplet. Because SDSS stellar spectra cover most of the entire optical
wavelength range, one can apply several approaches, using different wavelength
regions, in order to glean optimal information on stellar parameters. The
combination of multiple techniques results in estimates of stellar parameters
that are more robust over a much wider range of $T_{\rm eff}$, log $g$, and
[Fe/H] than those that might be produced by individual methods. In this first
paper of a series, we describe the SEGUE Stellar Parameter Pipeline (hereafter,
SSPP), which implements this ``multi-method'' philosophy. We also carry out a
number of tests to assess the range of stellar atmospheric parameter space over
which the estimates obtained by the SSPP remain valid. The second paper in the
SSPP series (Lee et al. 2007b; hereafter Paper II) seeks to validate the radial
velocities and stellar parameters determined by the SSPP by comparison with
member stars of Galactic three globular clusters (M~15, M~13, and M~2) and two
open clusters (NGC~2420 and M~67). A comparison with an analysis of
high-resolution spectra for SDSS-I/SEGUE stars is presented in the third paper
in this series (Allende Prieto et al. 2007; hereafter Paper III).

Section 2 describes determinations of radial velocity used by the SSPP. The
procedures used to obtain an estimate of the appropriate continuum, and the
determination of line indices, is explained in \S 3. The methods that the SSPP
employs for determining stellar parameters are presented in \S 4. Section 5
addresses the determinations of auxiliary estimates of effective temperature,
based on both theoretical and empirical approaches; these methods are used for
stars where adequate estimates of $T_{\rm eff}$ are not measured by our primary
techniques. A decision tree that gathers the optimal set of parameter estimates
based on the multiple methods is introduced in \S 6. Section 7 summarizes the
conditions for raising various flags to warn potential users where uncertainties
in parameter determinations remain. In \S 8, we discuss validation of the
parameters determined by the SSPP based on SDSS-I/SEGUE stars for which we have
obtained higher dispersion spectroscopy on various large telescopes, and also
compare the parameters with those of likely member stars of Galactic open and
globular clusters. Assignments of approximate MK spectral classifications are
described in \S 9. In \S 10 we describe several methods (still under testing)
for the determination of distance estimates used by the SSPP. Section 11
presents a summary and conclusions.

In the following, the colors ($u-g$, $g-r$, $r-i$, $i-z$, and $B-V$) and
magnitudes ($u,~g, ~r,~i,~z,~B,~\rm and~V$) are understood to be de-reddened and
corrected for absorption (using the dust maps of Schlegel et al. 1998), unless
stated specifically otherwise.




\section{Determinations of Radial Velocities}

\subsection{The Adopted Radial Velocities Used by the SSPP}

The {\it spZbest} fits file, which is generated from the SDSS spectroscopic
reduction pipeline, provides two estimated radial velocities. One is an
absorption-line redshift computed from a cross-correlation procedure using
templates that were obtained from SDSS commissioning spectra (Stoughton et al.
2002). Another estimate comes from performing a ``best-match'' procedure that
compares the observed spectra with externally measured templates (in this case,
the ELODIE library high-resolution spectra, as described by Prugniel $\&$
Soubiran 2001), degraded to match the resolving power of SDSS spectra.

Previous experience with the analysis of SDSS stellar spectra suggested that the
radial velocity estimated from the ELODIE template matches is often the best
available estimate, in the sense that it is the most repeatable, based on
spectra of ``quality assurance'' stars with multiple determinations. However,
there are some cases where the quoted error of an ELODIE spectral match velocity
is larger than expected; hence we also make use of the cross-correlation radial
velocity, in the following manner. If the velocity determined by comparison with
the ELODIE templates has a reported error of 20 km s$^{-1}$ or less, then this
velocity is adopted and the radial velocity flag is set to `RVOK(20)'. If the
error from the ELODIE template comparison is larger than 20 km s$^{-1}$ and the
relative difference between the two reported radial velocities is less than 40
km s$^{-1}$, then we take an average of the two techniques, and the radial
velocity flag is set to `RVOK(40)'. If the error in the reported ELODIE velocity
is larger than 20 km s$^{-1}$, and the difference of between the two estimates
is between 40 and 100 km s$^{-1}$, we take an average of the two and the radial
velocity flag is set to `RVOK(100)'.

If none of the above conditions are satisfied (which happens only rarely, and
mainly for quite low $S/N$ spectra, or for hot/cool stars without adequate
templates), then we obtain an independent estimate of the radial velocity based
on our own IDL routines. The calculation of the radial velocity is carried out
by determining wavelength shifts for several strong absorption line features (Ca
II K, Ca II H, H$\delta$, Ca I, H$\gamma$, H$\beta$, Na I, H$\alpha$, and Ca~II
triplet). After ignoring the calculated velocity above $+$500 km s$^{-1}$ or
below $-$500 km s$^{-1}$ from the individual lines (which are very often
spurious), we obtain a 3$\sigma$-clipped average of the remaining radial
velocities. If this computed average falls between $-$500 km s$^{-1}$ and +500
km s$^{-1}$, we take the calculated radial velocity as the adopted radial
velocity and set the radial velocity flag to `RVCALOK'. We have noticed that
certain types of stars, in particular cool stars with large carbon enhancements,
present a challenge for the radial velocity estimates obtained by the SSPP. We
have developed new carbon-enhanced templates, based on synthetic spectra, which
appear to return improved estimates. These will be implemented in the next major
update of the SSPP, which we anticipate applying for the next data release,
DR-7.

It should be noted that many of the techniques used for atmospheric parameter
estimation in the SSPP work well even when the velocity determination for a
given star has errors of up to 100 km s$^{-1}$ or more. The reason for this is
that many of our methods compare a wide spectral range with the synthetic
spectra, rather than a line-by-line comparison, to determine stellar parameters.
Even for those techniques that employ line-index approaches, the SSPP employs
relatively wide bandwidths, which will tend to mitigate small variations due to
radial velocity errors. Thus, small shifts in a spectrum due to a poor radial
velocity determination will not strongly influence our estimates of the stellar
parameters. Hence, we choose not to ignore spectra with high velocity errors,
but rather, indicate caution with the appropriate radial velocity flag.

If none of the above methods yields an acceptable estimate of radial velocity,
or if the reported velocity is apparently spurious (greater than 1000 km
s$^{-1}$ or less than $-1000$ km s$^{-1}$), we simply ignore the spectrum of the
star in our subsequent analysis, and set the radial velocity flag to `RVNOTOK'.

\subsection{Checks on Radial Velocities -- Zero Points and Scatter}

In order to check on the accuracy of the radial velocities adopted by the SSPP,
we compare with over 150 high-resolution spectra of SDSS-I/SEGUE stars that have
been obtained in order to calibrate and validate the stellar atmospheric
parameters. Table 1 summarizes the available high-resolution data. We
plan to continue enlarging this sample of validation/calibration stars in the
near future.

The high-resolution spectra have been reduced and analyzed independently by two
of the present authors (C. A. and T.S.). A detailed discussion is presented in
Paper III. During the course of deriving the stellar parameter estimates from
the high-resolution spectra, the radial velocities of stars are first measured.
Note that C.A. only considered the HET spectra, while S.T. considered all
available spectra. Thus, only the HET stars have radial velocities obtained by
both analysts; for these stars we take an average of their independent
determinations, which typically differ by no more than 1$-$2 km s$^{-1}$.
A more detailed discussion is presented in Paper III.

After rejecting problematic spectra (e.g., low $S/N$ high-resolution
spectra, or stars that appear to be spectroscopic binaries at high
spectral resolution), 125 stars remain to compare with the adopted
radial velocity results from the SSPP. Figure 1 shows the results of
these comparisons. A consistent offset of $-$6.94 km s$^{-1}$ (with
a standard deviation of 4.82 km s$^{-1}$) is computed from a
Gaussian fit to the residuals; this offset appears constant over the
color range $0.1 \leq g-r \leq 0.9$. An additional comparison with
the radial-velocity distribution of likely member stars in the
Galactic globular clusters M~15 and M~13 reveals similar offsets
($-$6.8 km s$^{-1}$ and $-$8.6 km s$^{-1}$, respectively; see Paper II). The
origin of this velocity offset might stem from different algorithms in the line
fits to arc and sky lines (Adelman-McCarthy et al. 2007b). It should be noted
that an offset of $+$7.3 km s$^{-1}$ is added to all DR-6 (Adelman-McCarthy et
al. 2007b) stellar radial velocities. This offset was computed by averaging the
offsets of $-$6.8 km s$^{-1}$(M~15) and $-$8.6 km s$^{-1}$(M~13), and the $-$6.6
km s$^{-1}$ offset that was obtained from a preliminary result of a
high-resolution spectroscopic analysis of SDSS-I/SEGUE stars that was carried
out prior to DR-6 (Adelman-McCarthy et al. 2007b). Now that a recent
high-resolution spectroscopic analysis of SDSS-I/SEGUE stars (Paper III)
indicates an offset of about $-$6.9 km s$^{-1}$, the average offset is +7.4 km
s$^{-1}$. Therefore, in future data releases (e.g, DR-7), this very minor
difference will likely be reflected in all adopted stellar radial velocities.
However, in order to account for its presence and to be consistent with the DR-6
database, we apply this empirical $+$7.3 km s$^{-1}$ shift to each adopted
radial velocity obtained by the SSPP. After application of this velocity shift,
the zero-point uncertainties in the corrected radial velocities determined by
the SSPP (and the SDSS spectroscopic reduction pipeline it depends on) should be
close to zero, with a random scatter on the order of 5 km s$^{-1}$ or less. Note
that the scatter in the determination of radial velocities, based on the average
displacements of the ``quality assurance'' stars with multiple measurements
varies from 5.0 km s$^{-1}$ to 9.0 km s$^{-1}$, depending on spectral type, and
exhibits a scatter of 2 km s$^{-1}$ between observations obtained with the
``faint'' and ``bright'' spectroscopic plug-plates (Adelman-McCarthy et al.
2007b).

\section{Calculation of Line Indices}

The initial step in calculation of line indices for SDSS spectra is to transform
the wavelength scale of the original SDSS spectrum over to an air-based (rather
than vacuum-based) wavelength scale, and to linearly rebin the spectrum to 1\,
{\AA} bins in the blue (3800$-$6000\,{\AA}), and 1.5\,{\AA} bins in the red
(6000$-$9000\,{\AA}). This slightly larger bin size is due to the degradation of
the resolution in the red regions of the spectra. Then, based on the adopted radial
velocity described above, the wavelength scale is shifted to zero rest
wavelength.

The SSPP measures line indices of 77 atomic and molecular lines. Line index
calculations are performed in two modes; one uses a global continuum fit over
the entire wavelength range (3850$-$9000\,{\AA}), while the other obtains local
continuum estimates around the line bands of interest. The choice between which
mode is used depends on the line depth and width of the feature under
consideration. Local continua are employed for the determinations of stellar
atmospheric parameters based on techniques that depend on individual line
indices. Other techniques, such as the neural network, spectral matching, and
auto-correlation methods, require wider wavelength ranges to be considered; for
these the global continuum is used. We make use of the errors in the fluxes
reported by the SDSS spectroscopic reduction pipeline to measure the uncertainty
in the line indices. Details of the procedures used to obtain the continuum fits
and line index measures (and their errors) are discussed below.

\subsection{Continuum Fit Techniques}
\subsubsection{Global Continuum Fit}

Determination of the appropriate continuum for a given spectrum is a delicate
task, even more so when it must be automated, and obtained for stars having wide
ranges of effective temperatures, as is the case for the present application.

In order to obtain a global continuum fit, the SSPP first proceeds by fitting
the stellar spectrum to a seventh-order polynomial. For wavelengths less than
6500\,{\AA}, if the observed flux is below that of the initial continuum fit in
each pixel, we add the difference between the observed flux and initial
continuum fit to the value of the continuum fit at each pixel. For wavelengths
larger than 6500\,{\AA}, we overwrite the observed flux with the fitted
continuum function (plus errors). This modified flux vector becomes the input
for the next continuum estimate, again obtained by fitting a seventh order
polynomial. This loop is iterated twice. The resultant continuum is considered
the first-pass estimate of the global continuum level.

The reason for setting the wavelength cut to 6500\,{\AA} in the above algorithm
is to guard against the effects of, e.g., low signal-to-noise data in the red
for warmer stars, the peculiar flux distribution of extremely cool stars, or
poor sky-line subtraction around the Ca~II triplet in the SDSS pipeline (in some
cases). We find that if we use the filtered flux for the first-pass continuum
estimate, as above, we are better able to derive reasonable results in the red
portion of the spectrum.

After obtaining the first-pass estimated continuum, we reject points in the
modified flux vector that are more than 3$\sigma$ above the fitted continuum,
where $\sigma$ is the estimated error in the fitting function. The reason that
we reject the points which are 3 $\sigma$ above the fitted line is to remove sky
lines in the Ca II triplet region, which may remain in the modified flux vector
even after obtaining the first-pass continuum estimate. Whether or not we reject
points that are 3$\sigma$ below the first-pass continuum, there is not much
difference in the final continuum determination. This is because, in the course
of deriving the first-pass continuum, most of strong absorption lines are taken
care of already. Following this step, a sixth-order polynomial fit to the
3$\sigma$ clipped flux vector is carried out in order to obtain a second-pass
continuum estimate. The final global continuum level is determined from a
weighted sum of the two estimated continua (40\% of the first-pass continuum
level estimate is added to 60\% of the second-pass continuum level estimate).
The upper panel of Figure 2 shows an example of a fitted global continuum
obtained in this manner. The bottom panel is the normalized spectrum, obtained
by division of the spectrum by the global continuum fit. It can be seen that a
reasonable continuum estimate is obtained even in the region of the Ca~II
triplet, where residuals from poor sky subtraction can sometimes be problematic.

\subsubsection{Local Continuum Fit}

To compute a local continuum over the line band of interest, we first calculate
a 5$\sigma$-clipped average of the fluxes over the (blue and red) sidebands
corresponding to each feature, as listed in Table 2. From an average of these
values a linear interpolation procedure is carried out over the central line
band. This linearly interpolated flux is then connected piecewise with the
average fluxes of the red and blue sidebands, and a robust fit is performed over
the entire region of the blue sideband + line band + red sideband to derive the
final local continuum estimate.

\subsection{Measurement of Line Indices}

Line indices (or equivalent widths) are calculated by integrating a continuum
normalized flux over the specified wavelength regions of each line band. Two
different measurements of line indices, obtained from the two different
continuum methods described above, are reported, even though the line-index
based methods for stellar parameter estimates only make use of the local
continuum-based indices. In order to avoid spurious values for the derived
indices, if a given index measurement is greater than 100\,{\AA}, or is
negative, we set the reported value to $-$9.999. No parameter estimates based on
that particular line index are estimated.

Table 2 lists the complete set of line indices made use of by the SSPP. Note
that, unlike the case for most of the features in this Table, the line indices
listed in rows 74 (TiO1), 75 (TiO2), 76 (TiO3), 77 (TiO4), 78 (TiO5), 79 (CaH1),
80 (CaH2), 81 (CaH3), 82 (CaOH), and 83 (H$\alpha$) are calculated following the
prescription given by the ``Hammer'' program (Covey et al. 2007). The line index
for Ca~I at 4227 \AA~ and the Mg~I$b$ and MgH features around 5170
\AA~ are computed following Morrison et al. (2003), so that they
might be used to estimate log $g$, as described in \S 4.4.

We follow the Cayrel (1988) procedure to compute an error for each line index
measurement. The uncertainty ($W_{error}$) in the index or measured equivalent
width is:

\begin{equation}
  W_{error} = \frac{1.6\times(resolution \times pixel~size)^{1/2}}{SNR}
\end{equation}

\noindent where $SNR$ is the signal-to-noise ratio in the local region
of the spectrum, the $resolution$ is taken to be $\sim$ 2.5\,{\AA}, and the {\it
pixel size} is set to 1\,{\AA} for the blue region (3800$-$6000 \AA) and 1.5
\AA~ for the red region (6000$-$9000 \AA). The noise spectrum provided by the
SDSS spectroscopic reduction procedure is used to compute the local $SNR$.

\section{Methodology}

The SSPP employs a number of methods for estimation of effective
temperature, surface gravity, and metallicity based on SDSS
spectroscopy and photometry. In this section the methods used in the
SSPP are summarized. Since many of the methods implemented in the
SSPP are already described by previously published papers, we will
address those techniques briefly, and refer the reader to individual
papers for detailed descriptions. The methods that are introduced
here for the first time are explained in more detail. Note that some
approaches derive all three atmospheric parameters simultaneously,
while others are specific to an individual parameter.

\subsection{Spectral Fitting With the {\tt k24} and {\tt ki13} Grids}

These two methods are based on identification of the parameters for
a model atmosphere that best matches the observed fluxes in a
selected wavelength interval, as described in detail by Allende
Prieto et al. (2006). Classical LTE line-blanketed model atmospheres
are used to compute a discrete grid of fluxes, and interpolation
allows sub-grid resolution. The search is performed using the
Nelder-Mead algorithm (Nelder \& Mead 1965).

The same grid described by Allende Prieto et al. (2006) is used by
the SSPP. We refer to this set of model fluxes as the {\tt k24}
grid. It includes a predicted broadband color index ($g-r$), as well
as normalized spectral fluxes in the region 4400$-$5500\,{\AA} at a
resolving power of $R = 1000$. Kurucz (1993) model atmospheres and
simplified continuum opacities are used to calculate synthetic
spectra. The broadband photometry was derived from the spectral
energy distributions provided by Kurucz (1993), passbands for point
sources determined by Strauss \& Gunn (2001), and an assumed airmass
of 1.3.

In addition to the {\tt k24} grid, a second grid, referred to as
{\tt ki13}, is now implemented in the SSPP. This second grid covers
the same spectral window as {\tt k24}, but no photometry is
considered. The use of only the derived normalized spectra
de-couples the results based on this grid from reddening and
photometric errors, although valuable information, mainly on the effective
temperature, is sacrificed.

There are several other differences between the {\tt k24} and {\tt
ki13} grids. The new grid includes molecular line opacities, with
the most relevant molecules in the range of interest being the CH
G-band near 4300\,{\AA} as well as the MgH band. In addition, the
{\tt ki13} grid takes advantage of a novel concept that allows for a
significant increase in the speed of the calculation of model
fluxes. The relevant opacities are not calculated for all depths in
all models, but instead are obtained on a temperature and density
grid, and later interpolated to the exact points in any given model
atmosphere (Koesterke et al. 2007). The opacity grid includes 4
points per decade in density and steps of 125~K in temperature. With
these choices, linear interpolation leads to errors in the derived
normalized fluxes smaller than 1\%.

Allende Prieto et al. (2006) made use of several libraries of
observed spectra and atmospheric parameters to study systematic and
random errors obtained from the {\tt k24} analysis. Even at infinite
signal-to-noise ratios, random errors appear significantly larger
than systematic errors, and amount to 3\% in $T_{\rm eff}$, 0.3 dex
in $\log g$, and 0.2 dex in [Fe/H]. This is most likely the result
of over-simplified model fluxes with a solar-scaled abundance
pattern (including an enhancement of the $\alpha$ elements at low
iron abundance), which is too limited to account for the chemical
spread in real stars. The new {\tt ki13} grid offers a significant
improvement in random errors, which at high signal-to-noise amount
to 2 \% in $T_{\rm eff}$, 0.2 dex in $\log g$, and 0.1 dex in
[Fe/H], but a less robust behavior (due to the lack of color
information) at low signal-to-noise ratios. Small systematic offsets
in {\tt ki13} detected from the analysis of the spectra in the
Elodie library (Prugniel \& Soubiran 2001) are corrected using
linear transformations.

As discussed in Allende Prieto et al. (2006), the {\tt k24} and {\tt
ki13} approaches perform best in the range 5000~K $\leq$ $T_{\rm
eff}$ $\leq$ 7000~K, which corresponds to $0.1 \leq g-r \leq 0.7$;
the SSPP restricts the adopted parameters from these methods to this
color range. These methods are designated in the SSPP according to
the following: the $T_{\rm eff}$, log $g$, and [Fe/H] estimated with
the {\tt k24} are referred to as T9, G7, and M8, respectively, while
these parameters estimated from the {\tt ki13} grid are referred to
as T10, G8, and M9, respectively.

\subsection{The Ca~II K and Auto-Correlation Function Methods}

These methods are based on the procedures outlined by Beers et al. (1999), where
the interested reader should look for more details. A brief summary follows.

The Ca~II K method makes use of a ``band-switched'' estimate of the
pseudo-equivalent of the Ca~II K line at 3933\,{\AA}, in combination with an
estimate of a broadband color, to obtain a prediction of the [Fe/H] for a given
star. The approach has been used for two decades during the course of the HK
(Beers, Preston, $\&$ Shectman 1985, 1992) and the Hamburg/ESO objective prism
surveys (Reimers $\&$ Wisotzki 1997; Christlieb 2003) for the determination of
metallicities of stars with available medium-resolution (2--3
\,{\AA} resolution, similar to the resolution of the SDSS spectra)
follow-up spectroscopy. The original calibration is based on high-resolution
abundance determinations (and $B - V$ colors) for a sample of $\sim $ 500 stars.
It is clear that one adopts the assumption that the calcium abundance tracks the
iron abundance in a monotonic fashion, which is almost always valid. In the
process of deriving the estimate of [Fe/H], the relationship between [Fe/H] and
[Ca/Fe] used is as follows:

\begin{equation}
[\rm Ca/{\rm Fe}] = \left\{
\begin{tabular}{cc}
                           0            & if [Fe/H] $\ge  0 $   \\
                      $-0.267 \times $ [Fe/H]   & if $-1.5 \le$ [Fe/H] $< 0$ \\
                            $+0.4$      & if  [Fe/H] $< -1.5$   \\
\end{tabular}
\right.
\label{law}
\end{equation}

This method has been shown to perform well over a wide range of metallicities,
in particular for stars with [Fe/H] $< -1.0$; external errors from the
calibration indicate that it has an intrinsic error no greater than 0.15--0.20
dex in the color range 0.3 $\leq$ $B-V$ $\leq$ 1.2. Above [Fe/H] = $-1.0$, and
in particular for cooler stars (below $T_{\rm eff}$ = 5000 K), the Ca~II K line
gradually begins to saturate. As a result, for cool, metal-rich stars, the
method will generally return an estimate of [Fe/H] that is on the order of 0.5
dex too low. This is ameliorated somewhat by empirical corrections that are
built into the program used to calculated this estimate, but it remains a source
of concern. It is important to recognize that for stars with very low
metallicities, and for warmer stars in particular, the Ca~II K line is one of
the few (in some cases only) metallic lines available in medium-resolution
spectra. Hence, this estimator plays an especially important role in such
situations.

Clearly, in the present application, a measurement of $B-V$ is not available.
Hence, we used the observed (or predicted, when necessary) $g-r$ color to
estimate the $B-V$ color. In order to accomplish this task, we made use of
several hundred stars with existing $B-V$ colors obtained during the course of
the HK and Hamburg/ESO surveys that happened to fall in the SDSS footprint, and
as a result, had available $g-r$ colors (note that only the fainter,
non-saturated stars could be used). These stars covered a variety of
metallicities, but in particular a large number of stars with [Fe/H] $< -1.0$
were included. An approximate transformation, suitable for low-metallicity
stars, was obtained by Zhao \& Newberg (2006); the transform $B-V$ = 0.187 +
0.916($g-r)$ was employed. We plan on deriving a new calibration of this method,
using $g-r$ colors directly, based on the large set of high-resolution
observations of stars that are being obtained at present. This will eliminate
the uncertainty inherent in the application of an approximate color
transformation.

Comparison of the metallicities obtained from this method with those derived
from high-resolution spectroscopy of SDSS-I/SEGUE stars, and for member stars of
open and globular clusters with known [Fe/H], indicates that [Fe/H] for stars
with $g-r$ $>$ 0.7 are consistently underestimated (due to saturation of the
Ca~II K line); we consider only metallicities determined for stars with $0.1
\leq g-r \leq 0.7$ as valid estimates from this approach.

The auto-correlation function technique was developed as an alternative method
for metallicity estimation which should perform well at higher metallicities,
where the Ca~II K technique is limited by saturation. As described in Beers et
al. (1999), and references therein, the method relies on an auto-correlation of
a given stellar spectrum, which generates a correlation peak whose strength is
proportional to the frequency and strength of weak metallic lines in a given
spectrum. The more such lines exist, the stronger the signal. This function has
been calibrated as a function of $B-V$ color; as before, a transformation from
$g-r$ to $B-V$ is used in order to obtain a prediction of [Fe/H].

The auto-correlation signal is expected to depend strongly on the
signal-to-noise of a given spectrum, growing with decreasing $S/N$. Essentially,
in the low $S/N$ limit, this function is responding to noise peaks rather than
to the presence of metallic features. This effect can be corrected for -- the
measured $S/N$ over the region in which the auto-correlation function is
calculated enters as a part of the calibration, and is effectively subtracted
off. We performed such a procedure for a set of SDSS stellar spectra with
parameters obtained from an early version of the SSPP. While the
auto-correlation function method exhibits a rather small star-to-star abundance
scatter when applied to the spectra of stars from open and globular clusters of
known metallicity (indicating that it is performing well), it suffers a
significant metallicity offset (0.5 to 1.0 dex too low), suggesting that the
initial calibration to the SSPP was suspect. Although we calculate the value of
this function during the execution of the SSPP, we do not employ it in the final
derived abundance estimates. In the near future, we expect to obtain a
re-calibration of this approach, based on the high-resolution spectroscopic
observations that have now been carried out. This re-calibration will also be
performed directly to $g-r$ colors.

The [Fe/H] estimates from these two approaches are referred to as M4 for the
Ca~II K method, and as M5 for the auto-correlation function technique.

\subsection{Calibration of a Ca~II Triplet Estimator of Metallicity}

The SDSS spectra extend to sufficiently red wavelengths to include the prominent
Ca~II triplet feature, which covers a spectral region 8400$-$8700\,{\AA}. These
lines are known to be sensitive to both luminosity (surface gravity) as well as
metallicity, so care must be exercised in their use as a metallicity indicator.

We have employed a line index that measures the integrated strength of these
lines, corrected for the presence of Paschen H lines, which also occur in this
wavelength interval. The line index definition, and the calculation of the
summed index, is as described by Cenarro et al. (2001a,b). In order to calibrate
this index for use with SDSS spectra, we have taken the library of some 700
spectra (and their listed atmospheric parameters) given by Cenarro et al. (see
http: //www.ucm.es/info/Astrof/ellipt/CATRIPLET.html), rebinned the spectra to
the SDSS spectral resolution, and calculated the corrected Ca II triplet index.
This index, along with their listed de-reddened value of the $B-V$ color, are
used as inputs to an artificial neural network procedure in order to predict the
estimated [Fe/H]. This procedure is able to reproduce the metallicity of the
Cenarro et al. stars to within $\pm 0.3$ dex over the temperature range 4000~K
to 8000~K, with some residual sensitivity to surface gravity.

After significant testing, it was decided that the SDSS spectra in the regions
of the Ca~II triplet suffered from too much noise (often due to poor sky
subtraction) in this region in order for this approach to be implemented in the
present SSPP. However, we are now mounting a new effort to better clean the
Ca~II triplet region of residual sky noise in order to see if this indicator can
be salvaged. The [Fe/H] estimated from this method is referred to as M6.

\subsection{Calibration of a Gravity Estimator Based on the Ca~I (4227\,{\AA}) and
Mg I~$b$ and MgH Features}

Among the prominent metallic species in stellar spectra, the two that are most
sensitive to surface gravity are the Ca~I line at 4227\,{\AA} and the Mg~I~$b$
and MgH features around 5170\,{\AA}. Both of these lines exhibit sensitivity to
metallicity as well. We have adopted the line index measurements and quoted
atmospheric estimates of [Fe/H] for the dwarfs and giants in the calibration
sample of Morrison et al. (2003), which were measured at a similar spectral
resolution to the SDSS (2.5$-$3.5\,{\AA}). Surface gravity estimates for the
stars involved in this calibration were obtained from the compilation of Cayrel
de Strobel (2001), while $B-V$ colors were obtained from the SIMBAD database.

These indices, along with their de-reddened $B-V$ colors and [Fe/H], are used as
inputs to an artificial neural network procedure in order to predict the
estimated surface gravity log $g$. This procedure indicates that the prediction
errors of the surface gravity, based on the Ca~I and MgH indices are on the
order of 0.35 dex and 0.30 dex, respectively.

As indicated by Morrison et al. (2003), these two methods are valid in the color
range corresponding to $0.4 \leq g-r \leq 0.9$. The gravity estimated from Ca~I
is referred to G4, while that obtained from the MgH feature is referred to as
G5.

\subsection{Parameters Obtained from the Wilhelm et al. (1999) Procedures}

These methods are based on the routines described by Wilhelm, Beers,
\& Gray (1999; WBG99), to which we refer the interested reader. Extensions
of these methods used in the SSPP are described below.

The procedures implemented in the SSPP are optimized for two
separate temperature ranges, one for the warmer stars ($g-r$ $\leq$
0.25), and one for the cooler stars with redder colors than this
limit. The stellar parameter determinations make use of comparisons
to theoretical $ugr$ colors and line parameters from synthetic
spectra, both generated from ATLAS9 model atmospheres (Kurucz 1993).
The synthetic spectra used in these procedures were generated using
the spectral synthesis routine SPECTRUM (Gray $\&$ Corbally 1994).

For the hotter stars, the observed normalized spectra are fit with a Voigt
profile to determine the Balmer-line equivalent widths and the $D_{0.2}$ (the
line width at 20\% below the local pseudo-continuum level) widths for H$\delta$,
H$\gamma$, and H$\beta$. The combination of Balmer-line equivalent widths,
$D_{0.2}$, and $u-g$ and $g-r$ colors are used to establish initial $T_{\rm
eff}$ and log~$g$ estimates, computed from functional trends in the theoretical
model parameters. For stars cooler than $T_{\rm eff}$ $\sim$ 8000 K, the surface
gravity is mainly determined by the $u-g$ color. For hotter stars the surface
gravity is primarily determined by the $D_{0.2}$ parameter. A metallicity
estimate is determined through the use of a combination of the equivalent width
of the Ca~II K line and a comparison to synthetic spectral regions that contain
other metallic lines. Once an initial abundance is established, the procedure is
iterated to convergence in all three stellar parameters.

For the cooler stars, only the $g-r$ color is used to establish an initial
estimate of $T_{\rm eff}$. For these stars, log~$g$ is determined from the $u-g$
color for stars as cool as $T_{\rm eff}$ = 5750 K. For stars cooler than this
limit, the strength of MgH is compared to synthetic spectra of similar $T_{\rm
eff}$ and [Fe/H] through the use of a band-strength indicator. The metal
abundance is determined by the combination of the Ca~II K line strength and a
minimum $\chi^{2}$ comparison to metallic-line regions in the spectra. The
procedure is then iterated to convergence.

For stars with $S/N$ $>$ 10/1, errors on the order of $\sigma(T_{\rm eff})$ = 225
K, $\sigma(\log~g)$ = 0.25 dex, and $\sigma([\rm Fe/
\rm H])$ = 0.3 dex can be achieved with this technique. The color
range of $g-r$ over which this approach is used for the SSPP is
$-0.2 \leq g -r \leq $ 0.8. The effective temperature, surface
gravity, and [Fe/H] estimated from this technique are referred to as
T8, G6, and M7, respectively.

\subsection{The Neural Network Approach}

The SSPP implements a flexible method of regression that provides a global
non-linear mapping between a set of inputs (the stellar spectrum ${\rm \bf
x}_i$) and a set of outputs (the stellar atmospheric parameters, ${\bf
s}=\{T_{\rm eff}, \log~g, {\rm [Fe/H]}\}$). This method has been described in
detail by Re Fiorentin et al. (2007), to which the interested reader is referred
for more details.

For the present application, it should be noted that we have chosen not to include
input photometry, although this certainly could be added if desired. We build an
RR regression model (which means we are training the approach based on real, as
opposed to synthetic, spectra) to parameterize real spectra. The training and
evaluation data sets are taken from a set of 38,731 stars from 140 SEGUE plates,
in directions of low reddening, which have had atmospheric parameters estimated
by a preliminary version of the SSPP. This step, which one might think of as
``internal training,'' is clearly not optimal, as one would ideally like an
independent basis for the training. This was not possible, until recently, due to
the absence of an adequate noise model for SDSS spectra. Such a model has now
been developed; we are in the process of testing and evaluation of its use in
combination with a new grid of synthetic spectra, and anticipate implementing it
in future versions of the neural network approach.

Figure 3 compares our model estimates with those from the early version of the
SSPP on the evaluation set. Overall we see good consistency, especially for
stars with $T_{\rm eff} < 8000$~K ($\log~T_{\rm eff} = 3.90$). Above this
effective temperature our models underestimate $\log~T_{\rm eff}$
relative to the SSPP. Most regression models such as ours are designed to
interpolate, rather than extrapolate. Extrapolation of the model to estimate
atmospheric parameters that are not spanned by the training set is relatively
unconstrained. Furthermore, the accuracy of the RR model is limited by the
accuracy of the target atmospheric parameters used in training, as well as their
consistency across the parameter space. In this case, the SSPP estimates are
combinations from several estimation models, each of which operates only over a
limited parameter range. Thus, the transition we see above 8000 K may indicate a
temperature region where one of the SSPP sub-models is dominating the SSPP
estimates, and this is not well-generalized by our model. From this comparison,
we find that the accuracies of our predictions (mean absolute errors) for each
parameter are $\sigma(T_{\rm eff})$ = 170~K, $\sigma(\log g) = 0.36$ dex, and
$\sigma({\rm [Fe/H]})$ = 0.19 dex.

The RR model has the advantage that exactly the same type of data are used in
the training and application phases, thus eliminating the issue of discrepancies
in the flux calibration or cosmic variance of the two samples. Of course, this
requires an independent estimation method (``basis parameterizer'') to
parameterize the training templates (which itself must use synthetic models at
some level). Our regression model then automates and -- more importantly --
generalizes this basis parameterizer. Indeed, the basis parameterizer may even
comprise multiple algorithms, perhaps operating over different parameters ranges
or used in a voting system to estimate atmospheric parameters. This is true in
the present case, where the basis parameterizer comes from a preliminary version
of the SSPP.

Experience with the behavior of the neural network approach on the SDSS-I/SEGUE
data indicated that this method tends to underestimate [Fe/H] below $T_{\rm
eff}$ = 5000 K and above $T_{\rm eff}$ = 7500 K, so we restrict its application
to this range of temperature, which corresponds to $0.1 \leq g-r \leq 0.7$.
Estimates of $T_{\rm eff}$, log $g$, and [Fe/H] obtained from the neural network
approach are referred to as T7, G3, and M3, respectively.

\subsection{The $\chi^{2}$ Minimization Technique Using the {\tt NGS1} and {\tt NGS2} Grids}

\subsubsection{Grids of Synthetic Spectra}

We have made use of Kurucz's NEWODF models (Castelli \& Kurucz 2003), which
employ solar relative abundances from Grevesse \& Sauval (1998), to generate
two sets of grids of synthetic spectra. The model atmospheres assume
plane-parallel line-blanketed model structures in one-dimensional local
thermodynamical equilibrium (LTE), and an enhancement of alpha-element
abundances by +0.4 dex for stars with [Fe/H] $\leq -0.5$. These new models
include H$_{2}$O opacities, an improved set of TiO lines, and no convective
overshoot (Castelli, Gratton, \& Kurucz 1997).

For production of the synthetic spectra we employed the {\tt turbospectrum}
synthesis code (Alvarez \& Plez 1998), with solar abundances by Asplund,
Grevesse \& Sauval (2005), which uses the treatment of line broadening described
by Barklem \& O'Mara (1998). The sources of atomic lines used by {\tt
turbospectrum} come from largely the VALD database (Kupka et al. 1999).
Linelists for the molecular species CH, CN, OH, TiO, and CaH are provided by B.
Plez (see Plez \& Cohen 2005, private communication), while the lines of NH,
MgH, and the C$_{2}$ molecules are adopted from the Kurucz line lists (see http:
//kurucz.harvard.edu/LINELISTS/LINESMOL/). The grid of the synthetic spectra
produced has resolutions of 0.01\,{\AA} or 0.005\,{\AA}, and spans from 3500~K
$\leq T_{\rm eff} \leq$ 10,000~K in steps of 250~K, 0.0 $\leq \log g \leq $ 5.0
in steps of 0.25 dex, and $-4.0 \leq \rm{[Fe/H]} \leq$ +0.5 in steps of 0.25
dex. These synthetic spectra are referred to as the {\tt NGS1} grid. After their
generation, these synthetic spectra were degraded to the SDSS resolution $R =
2000$, using a Gaussian convolution algorithm, then sampled into 1\,{\AA} per
pixel for application of the spectral matching technique described below.

A second grid of model atmospheres was constructed from the Kurucz ATLAS9 models
(Castelli \& Kurucz 2003), which do not employ alpha-element enhancements for
models with [Fe/H] $\leq -0.5$. The {\tt turbospectrum} synthesis code was again
used to generate the synthetic spectra. The synthetic spectra have a resolution
of 0.1\,{\AA}, and cover 4000~K $\leq T_{\rm eff} \leq$ 8000~K in steps of 250~K,
0.0 $\leq \log g \leq $ 5.0 in steps of 0.25 dex, and $-3.0 \leq \rm [Fe/H]
\leq$ +0.5 in steps of 0.25 dex. Ranges in [$\alpha$/Fe] were introduced for
spectral synthesis, over $-0.2 \leq [\alpha/{\rm Fe}] \leq +0.8$, in steps of
0.2 dex for each value of $T_{\rm eff}$, log $g$, and [Fe/H]. These synthetic
spectra are referred to as the {\tt NGS2} grid. After their generation, these
synthetic spectra are also smoothed to the resolution of the SDSS spectrographs.
The primary purpose of creating the {\tt NGS2} grid is to enable (future)
methods for the determination of [$\alpha$/Fe] for stars in the range 4000~K
$\leq T_{\rm eff} \leq$ 8000~K. However, since this is an independent grid, it
is also possible to obtain another set of predicted stellar atmospheric
parameters for the stars within this temperature range.

\subsubsection{Pre-processing Observed Spectra for the $\chi^{2}$ Minimization Technique}

The observed SDSS spectra are processed as described in \S 3 above. The blue
region of the spectrum contains most of the information required to constrain
the stellar parameters, but for cooler stars, the observed signal-to-noise ratio
peaks in the red region. As a compromise, and in order to speed up the analysis,
we only consider the spectral range 4500$-$5500\,{\AA}. The spectrum under
consideration is normalized after obtaining a pseudo continuum over the
4500$-$5500\,{\AA} range. The continuum fit is carried out in a similar fashion
to that described in \S 3, but lower (third and fourth) order polynomials are
employed, due to the shorter wavelength coverage. The synthetic spectra that are
used to match with the observed spectra are also normalized in the same fashion
over the same wavelength range.

\subsubsection{The Parameter Search Technique}

Foloowing the above steps, we then carry out a search for the best-fit model
parameters, i.e., those that minimize the difference between the observed flux
vector, $\emph{O}$, and the synthetic flux vector, $\emph{S}$, as functions of $T_{\rm
eff}$, log $g$, and [Fe/H] using a reduced $\chi^{2}$ criterion. That is,

  \begin{equation}
  \chi^{2}/DOF = \sum_{i=1}^{m+1}(O_i - S_i)^{2} /\sigma_i^{2},
  \end{equation}

\noindent where $\sigma_i$ is the error in flux in the $i$th pixel.

To reduce the number of model spectra that must be considered in the calculation
of the the reduced $\chi^{2}$ values, we first obtain an approximate effective
temperature based on a simple approximation. This procedure, which is referred
to as the Half Power Point (HPP) method (Wisotzki et al. 2000), obtains an
estimate of the wavelength at which the total integrated flux over a spectrum is
equal to half of the flux obtained over the entire wavelength region (in this
case we use 3900$-$8000\,{\AA}). Since the flux distribution for a given stellar
spectrum varies strongly with effective temperature, once we have
determined the HPP wavelength, we are able to obtain a reasonably accurate
estimate of effective temperature (or a broadband color such as $g-r$) by
comparing with the HPP wavelengths obtained from a grid of synthetic spectra.
The relation between effective temperature and HPP wavelength is established by
fitting a polynomial:

  \begin{equation}
  T_{\rm eff} = (25.63 - 114.51\rm HPP + 177.17\rm HPP^{2} -
  93.55\rm HPP^{3}) \times 10,000~{\rm K}\label{eq1},
  \end{equation}

\noindent where, HPP = wavelength/10,000.

We initially select synthetic spectra over a broad range around this predicted
effective temperature, within $\pm$1500~K. For example, if the HPP predicted
temperature of a star is 5000~K, we consider models between 3500~K and 6500~K.
As long as the observed spectrum doesn't have a grossly incorrect
spectrophotometric calibration, the estimated temperature will be well within
this range. We then obtain the reduced $\chi^{2}$ values between the observed
and the selected synthetic spectra over a 4500$-$5500\,{\AA} wavelength window.

Considering the distribution of reduced $\chi^{2}$ values as a function of
effective temperature only, we now have 399 points (21 different gravities and
19 metallicities) in each temperature grid. We then iteratively clip, in each
temperature grid, points that have larger values of reduced $\chi^{2}$ than the
median of the reduced $\chi^{2}$ values in each temperature grid. This iterative
procedure is performed five times, with the clipped values being those that are
higher than the median of all surviving values in each step. After this, only a
few points in each temperature step are left. Figure 4 shows one example of the
result of this procedure. We then fit the reduced $\chi^{2}$ distribution to a
fourth-order polynomial function as a function of temperature. At this point, we
clip off any points that lie farther than 2$\sigma$ below and 1$\sigma$ above
the fitted curve. In this manner, we are able to easily find the likely global
minimum, instead of becoming stuck in an spurious local minimum. The adopted
effective temperature is taken as the minimum of this fit (shown as the red line
in Figure 4).

Once the effective temperature is determined, we are able to narrow down the
model grids further. To obtain estimates of [Fe/H] and log $g$, we select models
within $\pm$550~K of the estimated $T_{\rm eff}$, and then exploit the same
iterative procedure as above. Note that the values of reduced $\chi^{2}$ respond
sensitively to small changes in $T_{\rm eff}$ and [Fe/H], allowing for their
optimal determinations. However, variations in the log $g$ values do not
strongly impact the reduced $\chi^{2}$, due to a shortage of gravity-sensitive
lines in the spectral window we examine. This leads to potentially large errors
in the estimated log $g$. These procedures are applied to both the {\tt NGS1}
and {\tt NGS2} grids.

Figure 5 shows two examples of synthetic spectra with parameters set to those
estimated by the procedure described above, over-plotted on the observed
spectral data. The $T_{\rm eff}$, log $g$, and [Fe/H] estimated from {\tt NGS1}
are referred to as T6, G2, and M2, respectively, while the log $g$ and [Fe/H]
estimated from {\tt NGS2} are referred to as G1 and M1, respectively. No
independent estimate of $T_{\rm eff}$ is obtained from the {\tt NGS2} grid, as
it is essentially degenerate with that determined from the {\tt NGS1} grid.

\subsubsection{Comparisons with Spectral Libraries and Analysis of
High-Resolution SDSS-I/SEGUE Stars}

In order to validate that the {\tt NGS1} and {\tt NGS2} grid approaches perform
well in determining stellar parameter estimates, we compare the results from
these techniques with literature values from two spectral libraries: ELODIE
(Prugniel $\&$ Soubiran 2001; Moultaka et al. 2004) and MILES
(S\'anchez-Bl\'azquez et al. 2006), and with those derived from analysis of the
SDSS-I/SEGUE stars with available high-resolution spectroscopy.

\subsubsubsection{Validation from the ELODIE and MILES Spectral Libraries}

The spectra in the ELODIE library were obtained with the ELODIE spectrograph at
the Observatoire de Haute-Provence 1.93 m telescope, and cover the spectral
region 4000$-$6800\,{\AA}. We employ 1969 spectra of 1390 stars with a resolving
power $R = $ 10,000, which are publicly available as part of the ELODIE 3
release (Moultaka et al. 2004). The spectra are first smoothed with a Gaussian
kernel to match the SDSS resolution. Most of the spectra have quite high $S/N$
ratios, and are accompanied with estimated stellar parameters from the
literature. Each spectrum (and parameter estimate) has a quality flag ranging
from 0 to 4, with 4 being best. In our comparison exercise, we only select stars
with 4000~K $\leq$ $T_{\rm eff}$ $\leq$ 10,000~K with a quality flag $\geq$ 1
for the spectra and all of the parameters.

Our examination indicates that these two approaches work best in the range
5000~K $\leq T_{\rm eff} \leq $ 8000~K. Comparison plots between the literature
values and the estimated parameters in this temperature range for 562 stars
among the ELODIE spectral library are shown in Figure 6. A Gaussian fit to the
residuals of each parameter reveals that the {\tt NGS1} estimate of $T_{\rm
eff}$ is higher by 86~K ($\sigma$ = 96~K), the surface gravity is larger by 0.10
dex ($\sigma$ = 0.24 dex ), and the metallicity is smaller by 0.17 dex ($\sigma$
= 0.14 dex), on average. For cooler stars, with 4000~K $\leq T_{\rm eff} \leq $
5000~K, we find that $<\Delta(T_{\rm eff})>$ and $<\Delta([\rm Fe/\rm H]) >$ are
177~K ($\sigma$ = 145~K) and 0.09 dex ($\sigma$ = 0.20 dex), respectively, which
are relatively small offsets and scatter, while the $<\Delta$(log~$g)>$ is 0.48
dex with $\sigma$ = 0.36 dex.

Because the spectra from the ELODIE library are of very high quality, one might
wonder how the parameter estimates would compare for the lower $S/N$ data
included among the SDSS-I/SEGUE stars. In order to test this, we inject Gaussian
noise into the ELODIE spectra to force them to $S/N$ = 50/1, 25/1, 12.5/1, and
6.25/1 per pixel at 5000\,{\AA}, respectively, degrade them to the SDSS
resolution, and apply the same procedures as above for estimation of the stellar
atmospheric parameters. These test spectra, and more detailed information on
noise models can be found in ftp://hebe.as.utexas.edu/pub/callende/sdssim/.
Table 3 shows the results of this exercise. Inspection of this Table shows that,
for $S/N \geq 12.5/1$, the shifts and scatter in the determinations of the
parameters remain acceptably small.

The MILES library includes 985 spectra obtained with the 2.5m INT and the IDS
spectrograph at La Palma. The wavelength coverage is 3530$-$7430\,{\AA}, and the
resolution is $\sim$ 2.3\,{\AA} (S\'anchez-Bl\'azquez et al. 2006). Because the
resolution of the spectra is similar to that of SDSS spectra, we make use of the
original MILES spectra in the analysis. After dropping the spectra with missing
parameters, and outside the temperature range 5000~K $\leq T_{\rm eff} \leq $
8000~K, 367 spectra remain. Figure 7 shows the comparison plots between the
selected literature values and the parameters estimated from {\tt NGS1}
procedure. It is clear from this Figure that offsets and scatter for the three
atmospheric parameters are very close to those obtained by comparison with the
ELODIE spectral library. For the cooler stars, with 4000~K $\leq T_{\rm eff}
\leq $ 5000~K, we obtain $<\Delta(T_{\rm eff})>$ = 179~K with $\sigma$ = 133~K,
$<\Delta([\rm Fe/\rm H]) >$ = 0.06 dex with $\sigma$ = 0.17 dex, and
$<\Delta$(log~$g)>$ = 0.58 dex with $\sigma$ = 0.45 dex.

Very similar behaviors were found for comparison of the {\tt NGS2} grid
technique as for the {\tt NGS1} grid technique. Table 3 summarizes the offsets
between the literature values and the estimated parameters for both synthetic
grid approaches.

We conclude from these comparisons that estimates of $T_{\rm eff}$ and [Fe/H]
over 4000~K $\leq T_{\rm eff} \leq$ $ 8000~K$ (corresponding to $0.0 \leq g-r
\leq 1.2$) should be acceptable. Surface gravity estimates from these techniques
are sufficiently accurate over 5000~K $\leq T_{\rm eff} \leq $8000~K
(corresponding to $0.0 \leq g-r \leq 0.7$) for both synthetic grid approaches.
Both approaches require that, in order to obtain useful parameter estimates, the
$S/N$ ratio of the spectra should be larger than $S/N = 12.5/1$; for the purpose
of the SSPP we conservatively adopt $S/N \geq 20/1$ in the color range $0.0 \leq
g-r \leq 0.4$, and $S/N \geq 10/1$ in the color range $0.4 \leq g-r \leq
1.2$.

\subsubsubsection{Validation from SDSS-I/SEGUE Stars with Available High-Resolution
Spectra}

As part of a long-term program to validate and improve estimates of stellar
atmospheric parameters determined by the SSPP, over the past two years
we have obtained higher resolution spectra for over 150 SDSS-I and SEGUE stars.
The targets cover a wide range of temperature and metallicity, but somewhat less
so in surface gravity. Existing ``holes'' in the parameter space will be given
high priority for future high-resolution campaigns. The data have been
independently reduced and analyzed by two authors (C.A. and T.S.). For a
detailed description of these analyses, the interested reader is referred to
Paper III. Among this sample, we selected 124 stars which have single-lined
spectra, $S/N$ $>40/1$ per pixel at 6000 \AA, and with well-determined stellar
parameters. For simplicity of our present comparison, we adopt the mean values
of stellar atmospheric parameters obtained by the independent analysis efforts.

Figure 8 shows a comparison between the parameters estimated from the {\tt NGS1}
grid approach and those determined from high-resolution analysis, over 5000~K
$\leq T_{\rm eff} \leq $ 8000~K. As summarized in Table 3, unlike for the
comparisons with the spectral libraries, the temperature and surface gravity
estimates obtained from {\tt NGS1} present negligible zero-point offsets, 5~K ($\sigma = $
137~K) and 0.00 dex ($\sigma$ = 0.30 dex) dex, respectively. The average
metallicity offset ($-0.12$ dex) and scatter (0.17 dex) are very close to those
obtained by comparison with the two libraries. Considering the results from
these three different comparisons, a small systematic offset in our derived
metallicities from this method may exist. However, we don't adjust any of these
offsets in the current SSPP. Adjustments will be considered for future
refinements of the SSPP, once a significant number of additional high-resolution
data for SDSS-I/SEGUE stars are obtained.

\section{Empirical and Theoretical Predictions of $g-r$ Color and $T_{\rm eff}$}

\subsection{Predictions of $g-r$ Color}

For a variety of reasons (e.g., nascent saturation, difficulties with
de-blending of sources, high reddening, etc.), the SDSS PHOTO pipeline (Lupton
et al. 2001) occasionally reports incorrect, or less-than-optimal estimates of
the broadband colors for a given target. Because several of the methods we
employ in the SSPP require a good measurement of (at least) the $g-r$ color, it
is useful to check if the reported $g-r$ color is commensurate with that
predicted from the flux-calibrated spectrum of the source, or with the strength
of spectral lines that correlate with effective temperature. This predicted
color is used to raise a cautionary flag for stars with possibly incorrect
reported colors, within some tolerance. We have developed three different
methods to predict $g-r$ color in the SSPP, as described below.

\subsubsection{Prediction of $g-r$ Color from the Half Power Point Method}

The first technique, the Half Power Point (HPP) method (Wisotzki et al. 2000), has
been described in \S 4.7 above, in connection with refining grid searches of
parameter space. Here we obtain an empirical calibration of the (de-reddened)
$g-r$ by fitting a functional relationship between the HPP
wavelength of spectra for stars with well-measured SDSS colors, and located in
regions of high Galactic latitude, where reddening is minimal. The best-fit
relationship is of the form:

\begin{equation}
  g-r = -3.354 + 4.318\rm HPP + 3.247\rm HPP^{2},
\end{equation}

\noindent where HPP = wavelength/10,000. The expected error in
prediction is about 0.08 magnitudes, over a broad range of color.

The predicted color obtained in this fashion is (obviously) also a way to
identify stellar spectra with poor spectrophotometric flux calibrations. If the
observed color reported by SDSS PHOTO is believed to be correct, and there
remains a difference with the color obtained from the above relationship, one
might be justifiably concerned about the quality of the spectrophotometric
correction that has been applied. Unresolved binaries, especially those
involving a red and a blue member, can also be identified by looking for
discrepancies between the observed and predicted $g-r$ colors.

\subsubsection{Prediction of $g-r$ Color from the H$\delta$ and H$\alpha$ Lines}

The strengths of the Balmer lines of hydrogen are also tightly correlated with
$g-r$ color over wide ranges of effective temperature. We have made use of the
line indices for H$\delta$ and H$\alpha$, as determined by the SSPP, to obtain
the following relationships:

\begin{equation}
  g-r = 0.469 - 0.058HD24 \label{eq3}
\end{equation}

\noindent
and

\begin{equation}
  g-r = 0.818 - 0.092HA24 \label{eq3},
\end{equation}

\noindent where $HD24$ and $HA24$ are the H$\delta$ and H$\alpha$ line indices
calculated over a 24\,{\AA} band centered on these lines. Note that since the
H$\alpha$ line is stronger, at a given color, than the H$\delta$ line, it can be
used to determine predictions of colors for cooler stars. The H$\alpha$ line is
also located in a region of the spectrum where one expects generally fewer
problems with contamination of the index from nearby metallic features.

\subsection{Predictions of $T_{\rm eff}$}

Effective temperatures predicted by the observed $g-r$ color, or through the
strength of the Balmer lines, are sufficiently accurate to be considered as
auxiliary estimators to those methods described in \S 4. We obtain two
theoretical and three empirical temperature estimates during execution of the
SSPP.

\subsubsection{Theoretical $T_{\rm eff}$ Estimates}

Two theoretical temperature estimates are based on grids of synthetic spectra
generated using the Kurucz models described above, and by consideration of
predicted colors from the Girardi et al. (2004) isochrones. For the temperatures
based on Kurucz models, we calculate an estimated $g-r$ color, adopting the SDSS
filter and instrumental response functions (Strauss $\&$ Gunn 2001), then fit a
fourth-order polynomial:

\begin{equation}
  T_{\rm eff} = 7792.22 - 6586.18(g-r) - 4637.23(g-r)^{2} -
  1994.29(g-r)^{3} - 386.24(g-r)^{4} \label{eq3}
\end{equation}

\noindent In deriving the above relationship, we take into account
stellar models with atmospheric parameters in the range $-2.0 \leq {\rm[Fe/H]}
\leq -0.5$ and $3.0 \leq \log g \leq 5.0$, where most SDSS-I/SEGUE stars are
found. Stars at the extrema of these ranges will have less than ideal estimates
of temperature due to the sensitivity of $g-r$ color to either metallicity,
surface gravity, or both. The effective temperature estimated from this relation
is referred to as T3.

For the temperature estimates based on the Girardi et al. isochrones, we assume
that the stars are all older than 10 Gyrs, are moderately metal poor
(i.e., have metallicites in the range $-1.5 \leq {\rm [Fe/H]} \leq -0.5$), and
are subgiants or main-sequence stars, which is true for a great majority of the
SDSS-I/SEGUE stars. The relationship below, based on a third-order polynomial,
is referred to as T4:

\begin{equation}
  T_{\rm eff} = 7590.26 - 6191.78(g-r) - 4270.92(g-r)^{2} - 1225.12(g-r)^{3}
\end{equation}

\subsubsection{Empirical $T_{\rm eff}$ Estimates}

Two of the three empirical temperature estimates we employ are derived from the
Balmer-line strengths, similar to the color estimates discussed above, but calibrated
to the effective temperature estimates obtained from the methods discussed in \S
4. The temperatures estimated from the $HA$24 and $HD$24 indices, via the simple
linear relationships below, are referred to as T1 and T2, respectively.

\begin{equation}
  T_{\rm eff} = 4133 + 371HA24
\end{equation}
\begin{equation}
  T_{\rm eff} = 5449 + 206HD24
\end{equation}

\noindent We restrict the regions over which the above relationships are applied
to 1.0\,{\AA} $\leq$ $HA24$ $\leq$ 12.0\,{\AA} and 1.0\,{\AA} $\leq$ $HD24$
$\leq$ 15.0\,{\AA}, respectively.

The final empirical temperature estimate comes from the relationship
between the effective temperature derived from a previous version of
the SSPP and the observed $g-r$ color (Ivezi\'c et al. 2007). The
temperature estimated from the relationship below is referred to as
T5:

\begin{equation}
  \log~T_{\rm eff} = 3.8820 - 0.3160(g-r) + 0.0488(g-r)^{2} + 0.0283(g-r)^{3}.
\end{equation}

\noindent The above temperature estimates are taken into account by the SSPP
provided that the color flag (see below) is not raised, and the expected
temperature is beyond the region where the primary estimates derived from the
techniques described in \S 4 apply. That is, they are used when the expected
temperature is outside the range 4500~K $\leq T_{\rm eff} \leq$ 7500~K.

\section{Flags Raised During Execution of the SSPP}

It is important that the SSPP be able to identify situations where the quoted
atmospheric parameters may be in doubt, or simply to make the user aware of 
possible anomalies that might apply to a given star. We have
designed a number of flags which serve this purpose.

There are two primary categories of flags -- critical flags and cautionary
flags. When a critical flag is raised, the SSPP is set to either ignore the
determinations of atmospheric parameters for a given star, or it is forced (in
the case of the color flag described below) to take steps that differ from
normal processing in an attempt to rescue this information. Obviously, even when
information is salvaged, the presence of a critical flag means the user must be
aware that special steps have been taken, and the reported estimated parameters
must be viewed with this knowledge in mind. The second category of flags are the
cautionary flags, which are provided for user consideration, but are not
necessarily cause for undue concern. Indeed, sometimes these flags are raised
when all is in fact OK, but the flag has been raised due to a peculiarity in the
spectrum that is relatively harmless, and which will not unduly influence
determination of atmospheric parameters. The user should nevertheless be aware
of the existence of these flags.

The flags are combined into a single set of four letters, the meanings of which
are summarized in Table 4, and described below in more detail. Four placeholders
are used in order to accommodate cases where more than one sort of flag is
raised.

The nominal condition for the four letter flag combination is `nnnn', which
indicates that the SSPP is satisfied that a given stellar spectrum (and its
reported $g-r$ colors) has passed all of the tests that have been performed, and
the stellar parameters should be considered well determined.

The first letter in this combination is set to one of 10 different values: `n',
`D', `d', `H', `h', `l', `E', `S', `V', and `N'.  Their explanations follow:

\begin{itemize}

\item `n':  The letter `n' indicates nominal.

\item `D': The letter `D' indicates that a comparison of the breadth of the H$\delta$
line at 20\% below its continuum, $D_{0.2}$, and the line depth below the
continuum, $R_c$, relative to their expected relationship for ``normal stars'',
provided below, does not apply.  The expected relationship is given by:

  \begin{equation}
  R_{\rm c} = -0.009503 + 0.027740D_{0.2} -0.000590D_{0.2}^{2} + 0.000006D_{0.2}^{3}
  \end{equation}

\noindent If $D_{0.2}$ is greater than 35.0\,{\AA}, and the predicted
$R_{\rm c}$ from the above relationship is less than the measured value,
then the star is most likely a white dwarf. This is a critical flag.

\item `d': This flag is raised if $D_{0.2}$ is less than 35.0\,{\AA}, and the
predicted $R_{\rm c}$ from above is less than the measured value. In
this case, the star is most likely a sdO or sdB star.  This is a
critical flag.

\item `H': This flag is raised when the estimated $T_{\rm eff}$ from the SSPP is
greater than 10000~K, and is meant to indicate a hot star.  This is
a critical flag.

\item `h': This flag is raised if the estimated $T_{\rm eff}$ from the
SSPP is greater than 8000~K, and either of the line indices of He~I
(at 4026.2\,{\AA}) or He~I (at 4471.7\,{\AA}) is greater than
1.0\,{\AA}. This indicates that the star is likely to be a hot star.
This is a critical flag.

\item `l': This flag is raised if the SSPP judges the star to have a high
likelihood of being a late-type star (generally late K, M, or later
spectral type), beyond the ability of the present pipeline to
determine acceptable atmospheric parameter estimates. The condition
used for raising the `l' flag is that the Na line (5892.9\,{\AA})
index, as measured over a 24\,{\AA} band centered on this feature,
is larger than 10\,{\AA}, and the $g-r$ color is greater than 0.80.
This is a critical flag.

\item `E': This flag is raised if significant emission lines are detected
in a spectrum.  This is a critical flag.

\item `S': This flag is raised if the spectrum (according to the header information)
is a night-sky spectrum.  This is a critical flag.

\item `V': This flag is raised when an adequate radial velocity could not be found for
a given spectrum.  This is a critical flag.

\item `N': This flag is raised if the spectrum is considered noisy at the extremes
of the wavelength range (e.g., around Ca~II K and the Ca~II triplet).  This is a
cautionary flag.

\end{itemize}

The flags that are used to fill out the remaining three positions of the four
letter flag combination are `C', `B', `G', or `g', as described below:

\begin{itemize}

\item `C': This flag is raised if the SSPP is concerned that the reported
$g-r$ color is incorrect. As mentioned above, we calculate three estimates of
predicted $g-r$ colors, based on $HA24$, $HD24$, or the Half Power Point method.
For each of these three predicted colors, we find the one which is closest to
the reported $g-r$ color based on the photometry. If the difference between the
reported color and the closest predicted color is larger than 0.2 magnitudes,
the color flag (C) is raised. The SSPP is set up to proceed with its
calculations of atmospheric parameters using the predicted $g-r$ color. This
flag is always found in the second position of the combination flag parameters.
This is a critical flag.

\item `B': This flag is raised if the SSPP is concerned that there exists
a strong mismatch between the strength of the predicted H$\alpha$ line index
$HA24$, based on the measured H$\delta$ line index, $HD24$. For the great
majority of ``normal'' stars, the predicted value of the H$\alpha$ line index is
found to be $HA24$ = 2.737 + 0.775$HD24$. For stars with significant $HA24$ and
$HD24$ measurements (which we take to mean that the values of these indices
exceed zero by more than 2$\sigma$, where $\sigma$ is the error in the measured
line index), if the difference between the predicted $HA24$ line index and the
measured $HA24$ index is larger than 2.5\,{\AA}, then the `B' flag is raised.
This flag is always found in the third position of the combination flag
parameters. This is a cautionary flag.

\item `G or g': This flag is raised if the SSPP suggests that the star may exhibit a
strong (`G') or mild (`g') CH G-band (around 4300\,{\AA}), relative
to expectation for ``normal'' stars. This flag is always found in
the fourth position of the combination flag parameters. This is a
cautionary flag.

\end{itemize}

\section{The SSPP Decision Tree for Final Parameter Estimation}

The SSPP uses multiple methods in order to obtain estimates of the atmospheric
parameters for each star over a very wide range in parameter space. Each
technique has limitations as to its ability to estimate each parameter, arising
from, e.g., the coverage of the grids of synthetic spectra, the methods used for
spectral matching, and their sensitivity to the $S/N$ of the spectrum, the range
in parameter space over which the particular calibration used for a given method
extends, etc.. Hence, it is necessary to specify a prescription for the inclusion
or exclusion of a given technique for the estimation of a given atmospheric
parameter. At present, this is accomplished by the assignment of a null (0,
meaning the parameter estimate is dropped), or unity (1, meaning the parameter
estimate is accepted) value to an indicator variable associated with each
parameter estimated by a given technique. In the future, we plan to devise an
improved weighting scheme for the combinations of the parameter estimates, once
the grid of high-resolution spectroscopic determinations of atmospheric
parameters is more completely filled out.

The $S/N$ ratio of a given spectrum also plays a role in the final decision as to the
estimate of a set of atmospheric parameters, and the techniques used (which
differ in their sensitivity to $S/N$). Table 5 lists the ranges of $g-r$ and
$S/N$ where each particular method is considered valid. Note that slightly
higher requirements on $S/N$ presently exist for the bluer stars (those with
$g-r < 0.3$) due to the inherent weakness of the metallic lines. We are
presently exploring whether this requirement can be relaxed in the cases of
several estimators. All derived parameters that fall outside of the color and
$S/N$ ranges listed in this table, for a given technique, are set to $T_{\rm
eff}$ = $-$9999, log $g$ = $-$9.999, and [Fe/H] = $-$9.999 by the SSPP.

Recall that in cases where the color flag `C' is raised, the predicted $g-r$
color determined by the procedures described above is used as an input (rather
than the reported color) for the techniques that require this information.

\subsection {Decisions on Effective Temperature Estimates}

There are five primary temperature estimates determined by the SSPP, and an
auxiliary set of five empirically and theoretically determined estimates. Note
that a few of the primary techniques extend to temperatures below 4500~K and
above 7500~K, although the accuracy obtained by these are lower than in the
interval 4500~K $ < T_{\rm eff} < 7500$~K. Thus, for stars with temperatures
outside of this interval, we also include the auxiliary temperature estimates
(in fact, just those that lie within 3$\sigma$ of the mean of the full auxiliary
set) in assembling the final average estimate of $T_{\rm eff}$.

In cases where the color flag `C' is raised, we ignore all temperatures that
rely on the reported $g-r$ color, and only consider those based on spectroscopy
alone (e.g., the spectral matching techniques). A robust average of the accepted
temperature estimates (those with indicator variables equal to 1) is taken for
the final adopted temperature. An internal robust estimate of the scatter around
this value is also obtained.

\subsection {Decisions on Surface Gravity Estimates}

There are eight methods used to estimate surface gravity by the SSPP.
Application of the limits on $g-r$ and $S/N$ eliminates a number of these
estimates, and a robust average of the accepted log $g$ estimates (those with
indicator variables equal to 1) is taken for the final adopted surface gravity.
An internal robust estimate of the scatter around this value is also calculated.

\subsection {Decisions on [Fe/H] Estimates}

Nine different methods are employed to determine [Fe/H] in the present SSPP. As
before, indicator variables of 1 or 0 are assigned to the result from each
method, according to whether or not it satisfies the range of validity listed in
Table 5. Note that two of the methods, the auto-correlation function technique
(M5) and the Ca~II triplet technique (M6) always have their indicator variables
set to 0 at present, until better calibrations for these estimators can be
obtained.

Three final values of [Fe/H] are determined from the assembled set of accepted
estimators. These are referred to as the biweight\footnote{The biweight family
of estimators smoothly diminish the effects of outliers on the resulting central
location (``mean'') and scale (``standard deviation''). See Beers, Flynn, \&
Gebhardt (1990) and references therein.} [Fe/H], the refined [Fe/H], and the
adopted [Fe/H], as described below.

The biweight [Fe/H] is simply a robust average of all accepted estimates of
metallicity with indicator variables equal to 1.

The refined [Fe/H] value is determined as follows. First, we select a small
region of the ({\tt NGS1}) grid where we find spectra that (globally) most
closely match the adopted $T_{\rm eff}$, the adopted log~$g$, and the [Fe/H]
from individual techniques with indicator variables equal to 1. As an example,
if there exist five estimates of [Fe/H] (with indicator variables of 1, and
using the adopted $T_{\rm eff}$ and adopted log~$g$), then five synthetic
spectra with the same temperature and gravity, but five different metallicities,
are selected. We then calculate reduced $\chi^{2}$ values in two restricted
regions of wavelength space in the observed spectrum, relative to the selected
synthetic spectra. The regions considered are the Ca~II region (3900$-$4000\,
{\AA}) and the region surrounding the MgH feature (5000$-$5500\,{\AA}). These
two regions are selected because they retain the most sensitivity to metallicity
for the extrema of the range of [Fe/H] and temperature encountered, i.e., for
the most metal-poor/warm or metal-rich/cool stars. We then seek the
best-matching synthetic spectrum over these regions alone, based on the minimum
of reduced $\chi^{2}$ for both regions, respectively. This match is carried out
independently for each region. Once the best matching synthetic spectrum (hence
[Fe/H]) among the five selected synthetic spectra is identified, we select the
accepted [Fe/H] estimates from the SSPP that lie within $\pm$ 0.15 dex of the
[Fe/H] of the best-matching synthetic spectrum in each region. These are then
averaged in order to obtain the best available metallicity estimate for each
region. At this point we have two averaged estimates of [Fe/H], which may be the
same (i.e., the same synthetic spectrum matches both regions equally well), or
different.

If the adopted estimate of $T_{\rm eff}$ $<$ 5300~K , the refined [Fe/H]
estimate is set to the average value obtained from the MgH region. The refined
value is also set to the metallicity average obtained from this region for the
case where 5300 $\leq$ $T_{\rm eff}$ $\leq$ 6500~K, and the biweight [Fe/H] is
$> -1.2$. This value is also used when $T_{\rm eff}$ $>$ 6500~K and the biweight
[Fe/H] is $\geq -$0.8.

The refined [Fe/H] is set to the average obtained for the Ca~II region when
5300~K $\leq$ $T_{\rm eff}$ $\leq$ 6500~K, and the biweight [Fe/H] is $\leq
-$1.2. The refined value is also set to the metallicity average obtained from
this region for the case where $T_{\rm eff}$ $>$ 6500~K and the biweight [Fe/H]
is $< -$0.8.

The adopted [Fe/H] is then set as follows. If the difference between the
biweight [Fe/H] and the refined [Fe/H] is less than 0.15 dex, the adopted
[Fe/H] is set equal to the biweight [Fe/H]. If the difference exceeds 0.15 dex,
the adopted [Fe/H] is set to the average value of the biweight [Fe/H] and the
refined [Fe/H] estimators.

\section{Validation of Final SSPP Parameter Estimates}

We do not yet have at our disposal a completely satisfactory set of external
spectral libraries, with suitable wavelength coverage and available atmospheric
parameter estimates, that extends over the full range of parameter space explored
by techniques employed by the SSPP. Hence, we are limited to comparison with the
sets of parameters obtained from analysis of the high-resolution spectra for
SDSS-I/SEGUE stars obtained to date, and with information available from the
literature for stars in Galactic open and globular clusters that have been
observed during the course of the SDSS. We discuss these comparisons below.

\subsection{Validation from High-Resolution Spectroscopy}

Table 1 summarizes the high-resolution data for SDSS-I/SEGUE stars obtained to
date. Although the stars in this Table cover most of the range explored by the
SSPP techniques, there remain gaps in this coverage that we hope to fill
in the near future.

As noted above, these data have been reduced and analyzed independently by two
of the authors (C.A. and T.S.), making use of different methodologies. Details
are discussed in Paper III. Tables 6, 7, and 8 summarize the systematic offsets
and scatter obtained for estimates of $T_{\rm eff}$, log $g$, and [Fe/H] from
each of the techniques used by the SSPP, relative to high-resolution analyses
carried out individually, and collectively. The differences in the numbers of
stars considered independently arises because T.S. (results shown as `HA2' in
the Tables) analyzed all available spectra, while C.A. (results shown as `HA1'
in the Tables) performed analysis only for those stars observed with the
Hobby-Eberly Telescope (HET). In the analysis `HA2', two different approaches
were employed. The first is to use a routine for optimizing minimum distance.
This method was employed for the HET and Keck-ESI spectra. The second is the
traditional high-resolution analysis approach, using Fe I and Fe II lines to constrain
$T_{\rm eff}$, log g$g$, [Fe/H], and microturbulence. This approach was applied
to the Keck-HIRES and Subaru-HDS data. More detailed explanations on the methods can
be found in Paper III. The Keck-ESI, Keck-HIRES, and Subaru-HDS spectra with
available parameters are defined as `OTHERS' in Paper III. It is noteworthy that
in Paper III only HET data analyzed by C.A. are used to derive the empirical
random errors of the SSPP. However, in this paper, we consider all available
data with estimated parameters. The rows labeled `MEAN' in Tables 6, 7, and 8
are the averaged results from HA1 and HA2 (for stars in common), supplemented
with stars from HA2 where HA1 results were not obtained.

Figures 9, 10, and 11 provide comparisons of the estimates of atmospheric
parameters for individual techniques used by the SSPP with those obtained from
the high-resolution analysis, for $T_{\rm eff}$, log $g$, and [Fe/H],
respectively. Note that comparisons are given even for the methods that are not
fully implemented by the SSPP at present (e.g., the auto-correlation function
and Ca~II triplet metallicity estimators).

Comparison of the estimated temperatures from the SSPP indicates an overall very
satisfactory result, although Table 6 reflects an interesting result for $T_{\rm
eff}$; estimates are mostly higher for HA1 and mostly lower for HA2. However,
as can also be noted from this Table, the final adopted value of effective
temperature from the SSPP exhibits a very small offset ($+$55~K), and a
one-sigma scatter of 123~K, both of which are encouragingly small. It is clear
from inspection of Figure 9 that additional high-resolution observations are
required of stars with both higher and lower temperatures than the present
sample. The distribution of the final adopted temperatures appears very well
correlated with that of the mean values from the high-resolution analyses in the
range of 4500~K $\leq$ $T_{\rm eff}$ $\leq$ 7500~K.

Table 7 and Figure 10 reveal that methods G3, G6, G7, and G8 exhibit the highest
offsets relative to the high-resolution analyses. The behavior of G3 (which
comes from the neural network approach described in \S4.6 above), is understood
because that network was originally trained on a preliminary version of the
SSPP, and hence it ``inherited'' whatever uncertainties existed in surface
gravity estimates at that time. It is not presently clear what reasons, other
than just the difficulty of extracting accurate estimates of log $g$, might
explain the large offsets of techniques G6, G7, and G8. Inspection of Figure 10
makes it clear that we could benefit from the inclusion of additional stars with
lower surface gravities. Nevertheless, the adopted values for surface gravity by
the SSPP are reasonably well distributed around the one-to-one correlation line.
The offset and one-sigma scatter in the final adopted estimate of log $g$ are
+0.04 dex and 0.25 dex, respectively, which is surprisingly good for this
difficult-to-estimate parameter.

It is clear from the comparison of the estimated metallicities from the SSPP in
Table 8 and Figure 11 why we exclude the estimates M5 (the auto-correlation
function method) and M6 (the Ca~II triplet method) for the time being. The M5
estimates exhibit large (low) offsets for stars at higher metallicites, and a
large overall scatter at lower metallicities (the latter is an expected
behavior). The M6 estimates are systematically offset (high) from the expected
correlation, and also have a larger scatter than desired. There is also a
tendency for M3, the estimate obtained from the neural network method, to
underestimate the metallicities of the more metal-rich stars. As mentioned
previously, this is understood to be the result of training this technique on a
previous version of the SSPP. We are in the process of re-calibrating all three
estimators, and expect to implement them in future versions of the SSPP.

Inspection of Figure 11 indicates that we could benefit from the addition of
more stars with intermediate metallicities, as well as for stars at the lowest
metallicities. The mean offset ($-0.04$ dex) and one-sigma scatter (0.21 dex) of
the residuals between the SSPP predictions of [Fe/H] and the high-resolution
analysis are quite encouraging, at least over the parameter space explored to
date.

It is obvious that there still exists a ``clustering effect'' in the low-metallicity
regime ([Fe/H] $\sim$ -2.0) in Figure 11. This stems from analysis of HET and
Keck-ESI data by T.S.. In the analysis of the HET and Keck-ESI spectra, the
wavelength region of $4800-5300$\, {\AA} is used for deriving the parameters.
This region includes the H$\beta$, Mg I$b$, Fe I , Fe II , Ca I,
and Cr I lines. The H$\beta$ region is sensitive to $T_{\rm eff}$, irrespective
of metallicity, and is independent of log $g$ for $T_{\rm eff}$ cooler than
6000 K; as a result $T_{\rm eff}$ estimates have less influence from [Fe/H] and log $g$.
However, log $g$ and [Fe/H] have similar effects on Mg I and other neutral lines.
At higher metallicities, the Mg I$b$ lines develop strong wings, and are sensitive
to log $g$. Therefore, one can de-couple the variation in $\chi^2$ due to
changes in log $g$ and [Fe/H] much better at higher metallicities. At low
metallicities the wings are not visible, increasing the
degeneracy between [Fe/H] and log $g$. 

In summary, based on the sets of parameter comparisons with the high-resolution
analysis, in the effective temperature range of 4500~K $\leq T_{\rm eff} \leq$
7500~K the SSPP is capable of producing estimates of the atmospheric parameters
for SDSS-I/SEGUE stars to precisions of $\sigma (T_{\rm eff})$, $\sigma (\log
g$), and $\sigma({\rm [Fe/H]})$ of 135~K, 0.25 dex, and 0.21 dex, respectively,
after adding systematic offsets quadratically. These uncertainties will be
slightly reduced if we take into account the error contribution from the high
resolution analysis as in Paper III (e.g., in a manner of quadratic subtraction)
. However, it should be kept in mind that the stars for which these comparisons
are carried out are among the very brightest observed with SDSS, and the overall
precision of parameter determination will decline for fainter stars. We are in
the process of quantifying the accuracies and precisions of atmospheric
parameter determinations as a function of $S/N$, and will report on these
results in due course.

Paper III takes a slightly different approach to derive empirical external
errors of the parameters determined by the SSPP. In this paper, final
external uncertainties are $\sigma(T_{\rm eff})$ = 130~K, $\sigma(\log g)$ =
0.21 dex, and $\sigma([\rm Fe/\rm H])$ = 0.11 dex.

\subsection{Validation from Galactic Open and Globular Clusters}

Galactic open and globular clusters provide nearly ideal testbeds for validation
of the stellar atmospheric parameters estimated by the SSPP. In most clusters,
it is expected that the member stars were born simultaneously out of well-mixed,
uniform-abundance gas at the same location in the Galaxy. Therefore, the member
stars should exhibit very similar elemental-abundance patterns. During the
course of SDSS-I and tests for SEGUE, we have secured photometric and
spectroscopic data for the clusters M~13, M~15, NGC~2420, and M~67, and can make
use of these clusters for validation of the atmospheric parameters obtained by
the SSPP. A more detailed description of this validation can be found in Paper
II, to which we refer the interested reader. Here, we briefly report on just the
results of the [Fe/H] comparisons as a function of $g-r$ color.

Figure 12 shows, from top to bottom, the SSPP estimated metallicities for likely
member stars of M~15, M~13, NGC~2402, and M~67 as a function of $g-r$ color. The
solid red line is the literature value of metallicity, reported by Harris (1996)
for M~15 and M~13, and Gratton (2000) for NGC~2420 and M~67, while the dashed
green line indicates the Gaussian mean SSPP [Fe/H] for the likely member stars
(solid dots). Inspection of this Figure indicates that the SSPP obtains results
for M~15 that are about 0.14 dex higher than the literature value, agrees quite
well with the intermediate-metallicity clusters, and underestimates the
metallicity of M~67 by about 0.3 dex. In addition, their appears to exist a
slight trend of declining [Fe/H] with respect to $g-r$ in these determinations,
at least for the more metal-rich clusters. Observations of additional clusters,
especially of intermediate and near-solar metallicities, will clearly be
helpful.

\section{Assignment of Spectral Classifications for Early and Late-Type Stars}

It is often useful to group stars into rough MK spectral classifications. It
should be kept in mind, however, that for this, and any other exercise of
assigning MK spectral types, that the MK system {\it does not apply} to stars
other than Population~I. That is, typing of metal-poor Population II stars is,
{\it by definition}, not a strictly valid procedure. Nevertheless, the SSPP
attempts to carry out this exercise using two approaches. The first is based on
the spectral type listed in the ELODIE database for the best template match
obtained for the determination of radial velocity (as described above), and
applies to stars with spectral classes O to M.

For the coolest stars, measurement of accurate values of $T_{\rm eff}$, $\log
g$, and [Fe/H] from spectra dominated by broad molecular features becomes
extremely difficult (e.g., Woolf \& Wallerstein 2006). As a result, the SSPP
does not estimate atmospheric parameters for stars with $T_{\rm eff} <$ 4500 K,
but instead estimates the MK spectral type of each star using the ``Hammer''
spectral-typing software developed and described by Covey et al. (2007)
\footnote{The Hammer has been made available for community use: the IDL code can
be downloaded from \url{http://www.cfa.harvard.edu/~kcovey/}}. The Hammer code
measures 23 spectral indices, including atomic lines (H, Ca I, Ca II, Fe I, Mg
I, Na I) and molecular bandheads (CN, G band, TiO, VO, CaH, FeH), as well as a
select set of broadband color ratios. The best-fit spectral type of each target
is assigned by comparison to the grid of indices measured from more than 1000
spectral-type standards derived from spectral libraries of comparable resolution
and coverage (Allen \& Strom 1995; Prugniel \& Soubiran 2001; Hawley et al.
2002; Bagnulo et al. 2003; Le Borgne et al. 2003; Valdes et al. 2004;
Sa\'nchez-Bl\'azquez et al. 2006).

Tests of the accuracy of the Hammer code with degraded ($S/N \sim\ 5/1$) STELIB
(Le Borgne et al. 2003), MILES (S\'anchez-Bl\'azquez et al. 2006), and SDSS
(Hawley et al. 2002) dwarf template spectra reveal that the Hammer code assigns
spectral types accurate to within $\pm$ 2 subtypes for K and M stars. The Hammer
code also returns results for warmer stars, but as the set of indices used is
optimized for cool stars, typical uncertainties are $\pm 4$ subtypes for A-G
stars at $S/N \sim$ 5/1; in this temperature regime, the SSPP atmospheric parameters
are a more reliable indicator of $T_{\rm eff}$.

Given the science goals of, in particular, the SEGUE program, we emphasize two
limitations to the accuracy of spectral types derived by the Hammer code:

\begin{itemize}

\item The Hammer code uses spectral indices derived from dwarf standards; spectral types
assigned to giant stars will likely have larger, and systematic, uncertainties.

\item The Hammer code was developed in the context of SDSS-I's high Galactic latitude
spectroscopic program; the use of broadband color ratios in the indices 
will likely make the spectral types estimated by the Hammer code particularly
sensitive to reddening. Spectral types derived in areas of high extinction (i.e.,
low-latitude SEGUE plates) should be considered highly uncertain until verified
with reddening-insensitive spectral indices.

\end{itemize}

\section{Preliminary Distance Estimates}

A number of techniques are presently being explored by members of the SEGUE team
in order to derive the best available estimates of distances for stars in the
SDSS/SEGUE database. Many rely on the existence of either theoretical or
empirical transformations of the substantial amount of photometric data
that exists for Galactic clusters obtained with photometric systems other than
$ugriz$. These will be reported on in due course (Morrison et al. 2007, in
preparation). For now, the SSPP assigns preliminary distance estimates for stars
of different luminosity classifications based on the empirical fits of Beers et
al. (2000) to the observed color-magnitude diagrams of Galactic clusters of
different metallicities and with reasonably well-known distances (in the Johnson
$V,~B-V$ system). For convenience, we use the same transformations as mentioned
above, based on the work of Zhao \& Newberg (2006); $V = g - 0.561(g-r) -
0.004$, and $B-V$ = 0.187 + 0.916($g-r)$.

Beers et al. (2000) argue that their distances should be accurate to on the order of 10
$-$ 20\%; a typical value of 15\% can be adopted for our distance estimates,
although this needs to be confirmed with future work.

The SSPP does {\it not} make a stellar luminosity classification, but rather, it
provides the atmospheric parameters from which
the user can make an appropriate choice. Distance estimates are obtained for the
following rough luminosity classes: Dwarf, Main-Sequence Turnoff, Giant,
Asymptotic Giant Branch, and Field Horizontal-Branch. Note that distance
estimates are obtained for all (feasible) cases where a star may fall into one
or more of these classifications, but only one of the listed distances is likely
to apply to a given star. The choice is up to the user.

Two other methods for distance estimates are obtained by the SSPP. The first is
described by Allende Prieto et al. (2006), to which the interested reader is
referred for a detailed description. The second is based on the isochrones in
the $ugriz$ system developed by Girardi et al. (2004). Our initial tests of this
method did not converge as well as was hoped. Although we will continue to work
on this, and other methods, for distance estimates, we do not fully implement
the Girardi et al. isochrone approach in the SSPP for the DR6 (Adelman-McCarthy
et al. 2007b) release.

\section{Conclusions}

We have described the development and execution of the SEGUE Stellar Parameter
Pipeline (SSPP), which makes use of multiple approaches in order to estimate the
fundamental stellar atmospheric parameters (effective temperature, $T_{\rm
eff}$, surface gravity, log $g$, and metallicity, parameterized by [Fe/H]) for
stars with spectra and photometry obtained during the course of the original
Sloan Digital Sky Survey (SDSS-I) and its current extension (SDSS-II/SEGUE).

The use of multiple approaches allows for an empirical determination of the
internal errors for each derived parameter, based on the range of the reported
values from each method. From consideration of about 140,000 spectra of stars
obtained during SDSS-I and SEGUE that have derived stellar parameters available
in the range 4500~K $\le$ $T_{\rm eff}$ $\le$ 7500~K, typical internal errors
obtained by the SSPP are $\sigma(T_{\rm eff})$ = 73~K (s.e.m), $\sigma(\log~g)$
= 0.19 (s.e.m), and $\sigma([\rm Fe/\rm H])$ = 0.10 (s.e.m). Paper III points
out that the internal scatter estimates obtained from averaging
the multiple estimates of the parameters produced by the SSPP underestimate
the external errors, owing to the fact that several methods in the SSPP use similar
same parameter indicators and atmospheric models.

The results of a comparison with an average of two different high-resolution
spectroscopic analyses of 124 SDSS-I/SEGUE stars suggests that the SSPP is able
to determine $T_{\rm eff}$, log $g$, and [Fe/H] to precisions of 135~K, 0.26
dex, and 0.21 dex, respectively, after combining small systematic offsets
quadratically for stars with 4500~K $\leq T_{\rm eff} \leq$ 7500~K. These errors
differ slightly from the those obtained by Paper III ($\sigma(T_{\rm eff})$ =
130~K, $\sigma(\log g)$ = 0.21 dex, and $\sigma([\rm Fe/\rm H])$ = 0.11 dex),
even though they share a common set of high-resolution calibration observations.
This arises because Paper III derived the external uncertainties of the SSPP
only taking into account the stars observed with the HET (on the grounds of
internal consistency). The sample referred to as OTHERS in Paper III exhibits
somewhat larger scatter in its parameters, when compared with those determined
by the SSPP. Observation of several hundred additional stars from SDSS-I/SEGUE
with HET is now underway. Thus, in the future, we will be able to use a
homogeneous sample gathered by HET in our tests. Also, additional
high-resolution data for stars outside of our adopted temperature range will
enable tests for both cooler and warmer stars.

Considering the internal scatter from the multiple approaches and the external
uncertainty from the comparisons with the high-resolution analysis together, the
typical uncertainty in the stellar parameters delivered by the SSPP are
$\sigma(T_{\rm eff})$ = 154~K, $\sigma(\log~g)$ = 0.31 dex, and $\sigma([\rm
Fe/\rm H])$ = 0.23 dex, over the temperature range 4500~K $\leq T_{\rm eff}
\leq$ 7500~K.

However, it should be kept in mind that the errors stated above apply for the
very highest $S/N$ spectra obtained from SDSS ($S/N >$ 50/1), as only quite
bright stars were targeted for high-resolution observations. In addition,
outside of the quoted temperature range (4500~K $\le$ $T_{\rm eff}$ $\le$
7500~K), we presently do not have sufficient high-resolution spectra to fully
test the parameters obtained by the SSPP.

The results of a comparison with likely member stars of a sample of Galactic
open and globular clusters suggest that SSPP may slightly overestimate [Fe/H]
(by $\sim 0.15$ dex) for stars with [Fe/H] $< -$2.0, and underestimate [Fe/H]
(by $\sim 0.30$ dex) for stars with near-solar metallicities. Slight trends of
[Fe/H] with $g-r$ are noticed for the higher metallicity clusters as well,
although further data will be needed in order to verify this.

Approximate spectral types are assigned for stars, based on two methods, with
differing limitations. A preliminary set of distance determinations for each
star is also obtained, although future work will be required in order to
identify the optimal method.

We conclude that the SSPP determines sufficiently accurate and
precise radial velocities and atmospheric parameter estimates, at
least for stars in the effective temperature range from 4500~K to
7500~K, to enable detailed explorations of the chemical compositions
and kinematics of the thick-disk and halo populations of the Galaxy.

\acknowledgements

Funding for the SDSS and SDSS-II has been provided by the Alfred P. Sloan
Foundation, the Participating Institutions, the National Science Foundation, the
U.S. Department of Energy, the National Aeronautics and Space Administration,
the Japanese Monbukagakusho, the Max Planck Society, and the Higher Education
Funding Council for England. The SDSS Web Site is http://www.sdss.org/.

The SDSS is managed by the Astrophysical Research Consortium for the
Participating Institutions. The Participating Institutions are the American
Museum of Natural History, Astrophysical Institute Potsdam, University of Basel,
University of Cambridge, Case Western Reserve University, University of Chicago,
Drexel University, Fermilab, the Institute for Advanced Study, the Japan
Participation Group, Johns Hopkins University, the Joint Institute for Nuclear
Astrophysics, the Kavli Institute for Particle Astrophysics and Cosmology, the
Korean Scientist Group, the Chinese Academy of Sciences (LAMOST), Los Alamos
National Laboratory, the Max-Planck-Institute for Astronomy (MPIA), the
Max-Planck-Institute for Astrophysics (MPA), New Mexico State University, Ohio
State University, University of Pittsburgh, University of Portsmouth, Princeton
University, the United States Naval Observatory, and the University of
Washington.

Y.S.L., T.C.B., and T.S. acknowledge partial funding of this work
from grant PHY 02-16783: Physics Frontiers Center / Joint Institute
for Nuclear Astrophysics (JINA), awarded by the U.S. National
Science Foundation. NASA grants (NAG5-13057, NAG5-13147) to C.A.P.
are thankfully acknowledged. J.E.N acknowledges support from
Australian Research Council Grant DP0663562. C.B.J and P.R.F
acknowledge support from the Deutsche Forschungsgemeinschaft (DFG)
grant BA2163.

\clearpage

\begin{figure}
\centering
\includegraphics[scale=0.6]{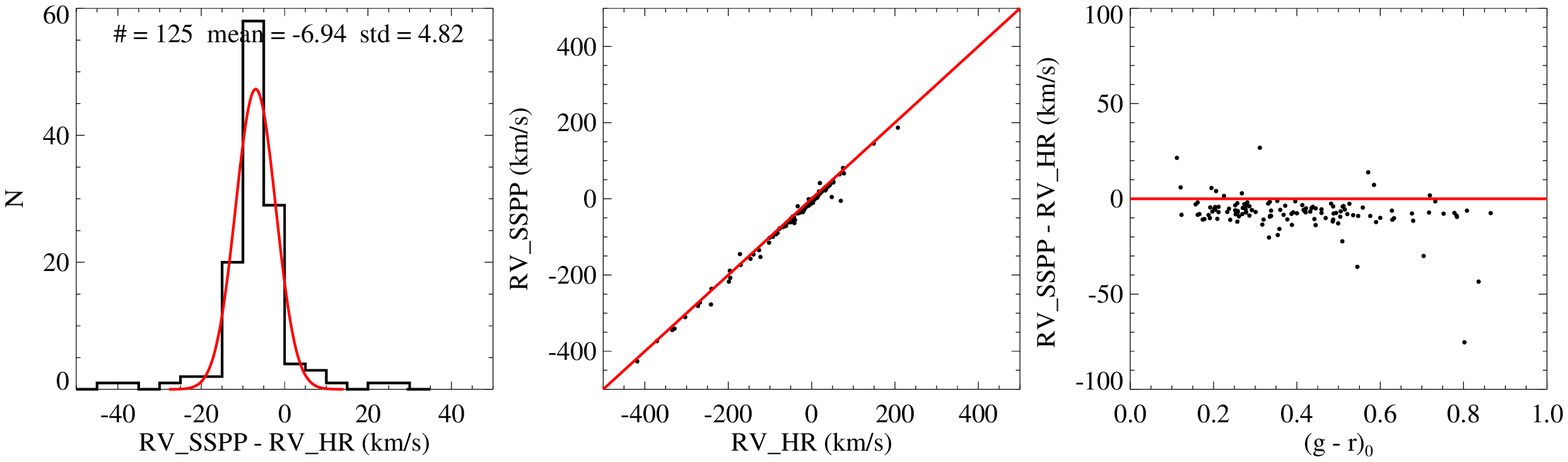}
\caption{A comparison plot of the radial velocity adopted by the
SSPP with that measured by the high-resolution analyses. There were
three different measurements for stars observed with HET for the
high-resolution data. A simple average of the three measurement was
taken for this comparison. An offset of $-$6.9 km s$^{-1}$ is
noticed from the Gaussian fit to the residuals. This offset appears
constant with $g-r$ in the right-hand panel.}
\end{figure}
\clearpage

\begin{figure}
\centering
\includegraphics[scale=0.8]{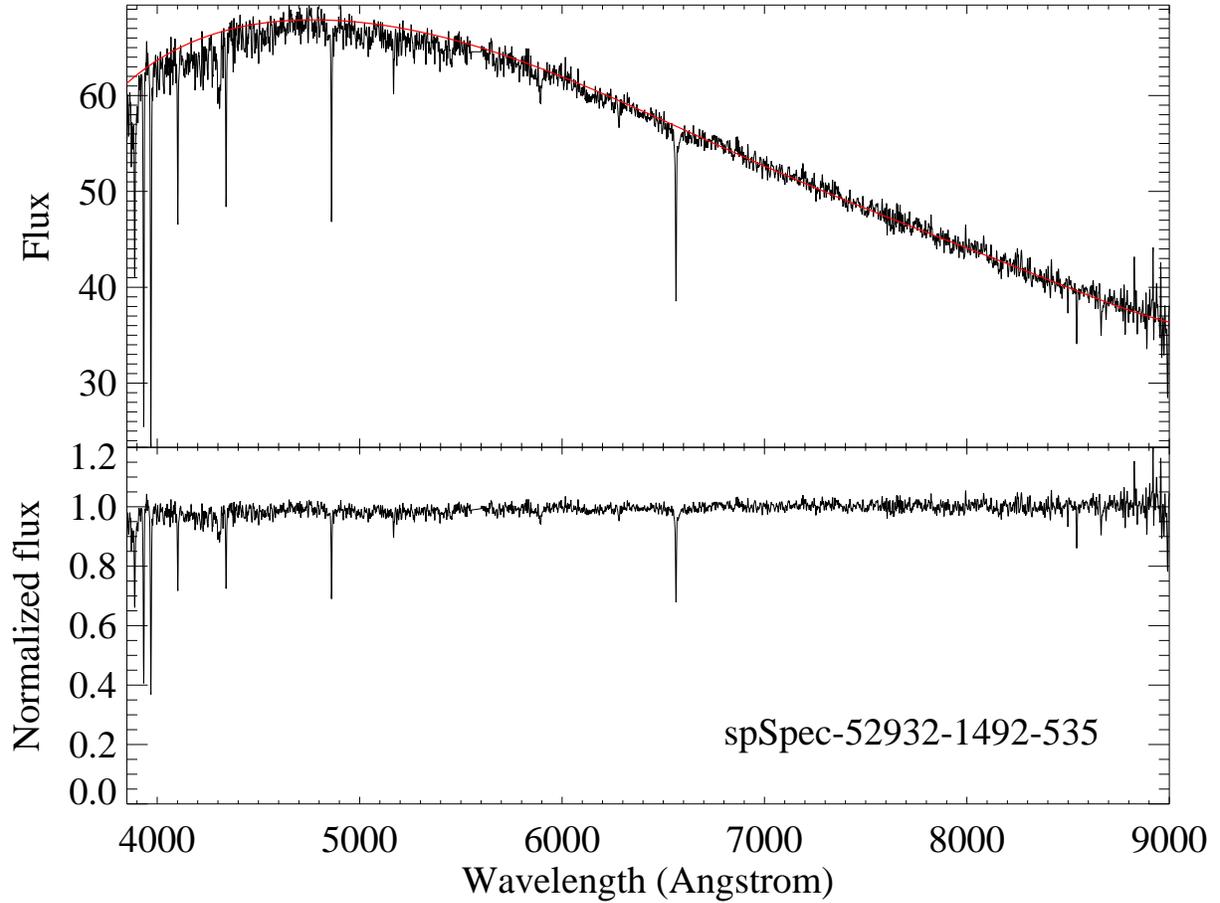}
\caption{An example of a fitted global continuum. The red line in
the upper panel is the fitted continuum over the 3850$-$9000 \,{\AA}
wavelength range; the black line is the observed spectrum. The bottom
panel shows the normalized flux.}
\end{figure}
\clearpage

\begin{figure}
\centering
\includegraphics[angle=-90,scale=0.6]{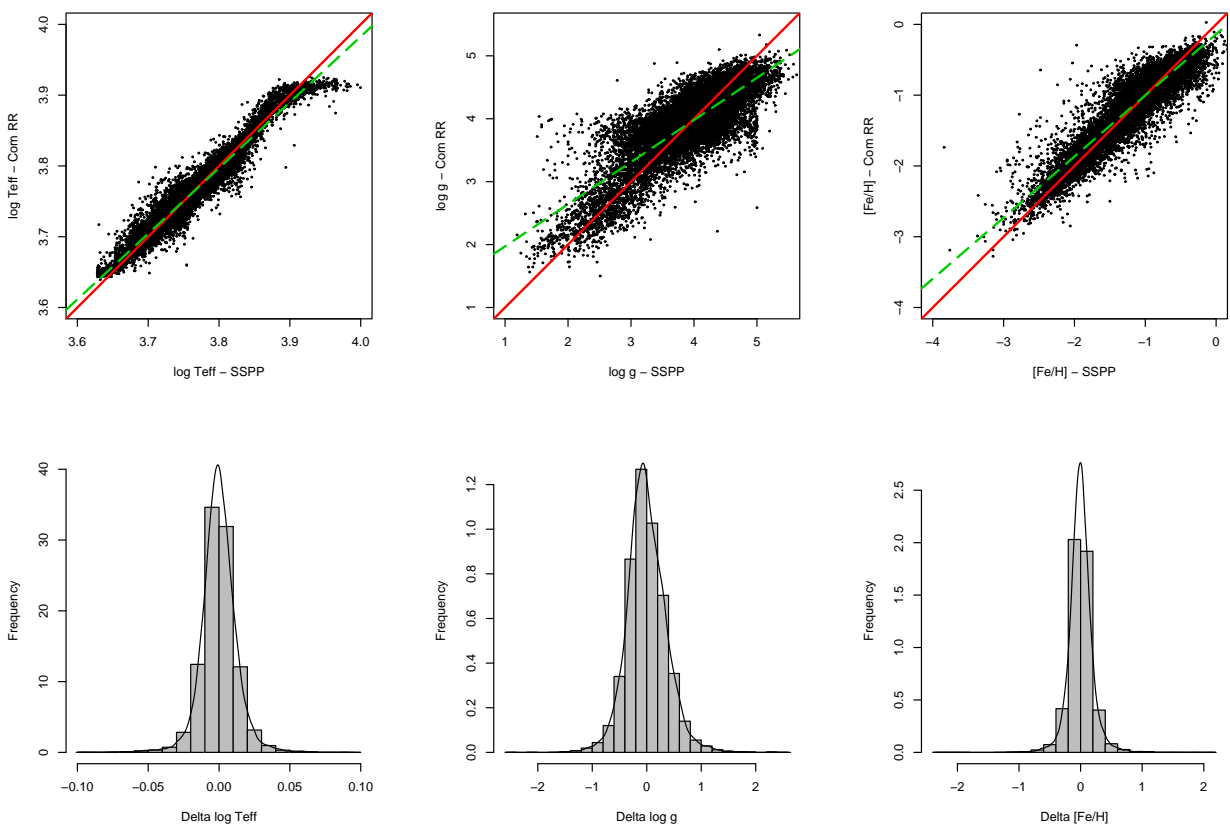}
\caption{Atmospheric parameter estimation with the RR model. We
compare our estimated log $T_{\rm eff}$, log $g$, and [Fe/H] with
those from a preliminary version of the SSPP on the 19,000 stars in
the evaluation set. The perfect correlation and a linear fit to the
data are shown with the solid and dashed lines respectively. The
histogram of the discrepancies (our estimates minus SSPP estimates)
are shown in the lower panels.}
\end{figure}
\clearpage

\begin{figure}
\centering
\includegraphics[scale=0.8]{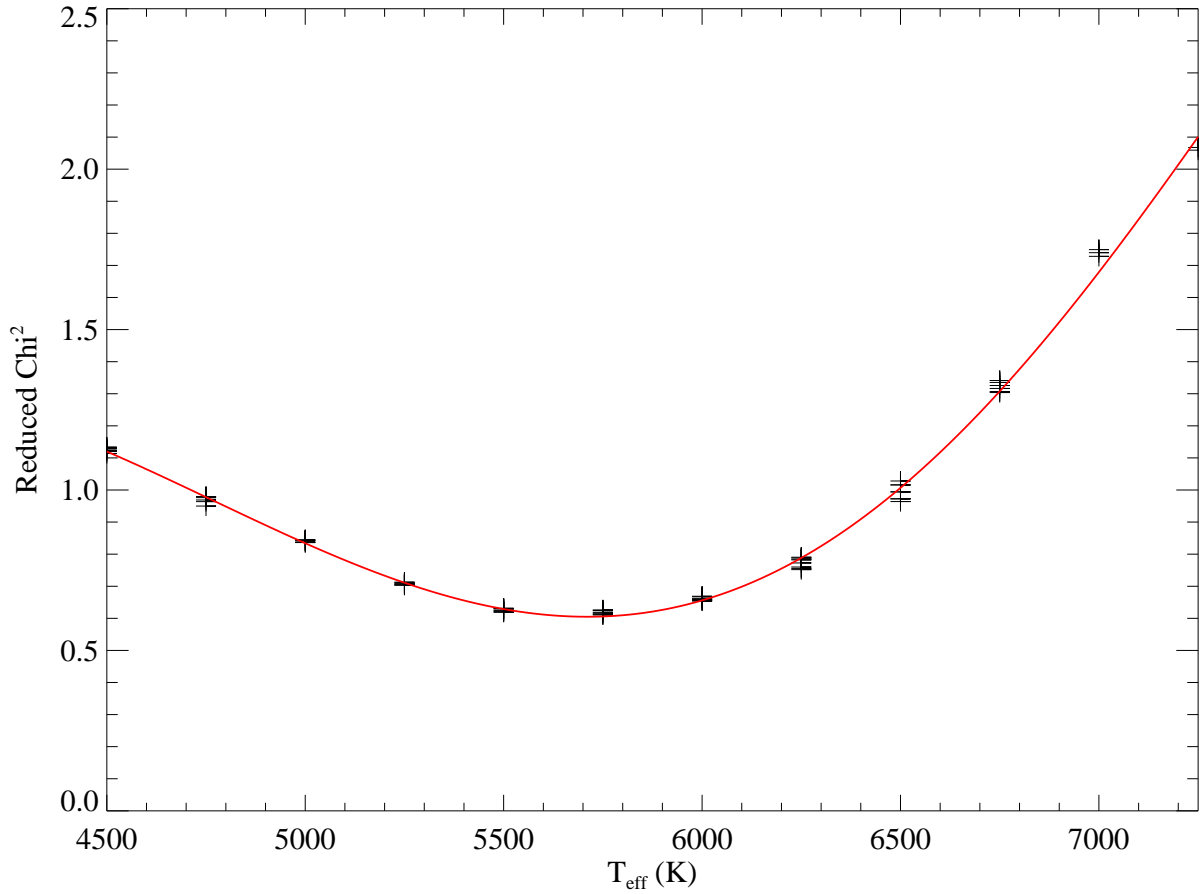}
\caption{An example of finding a minimum reduced $\chi^{2}$ value as
a function of $T_{\rm eff}$. The crosses are the remaining
values after application of an iterative rejection scheme. The red
curve is the final fitted function.}
\end{figure}
\clearpage

\begin{figure}
\centering \plottwo{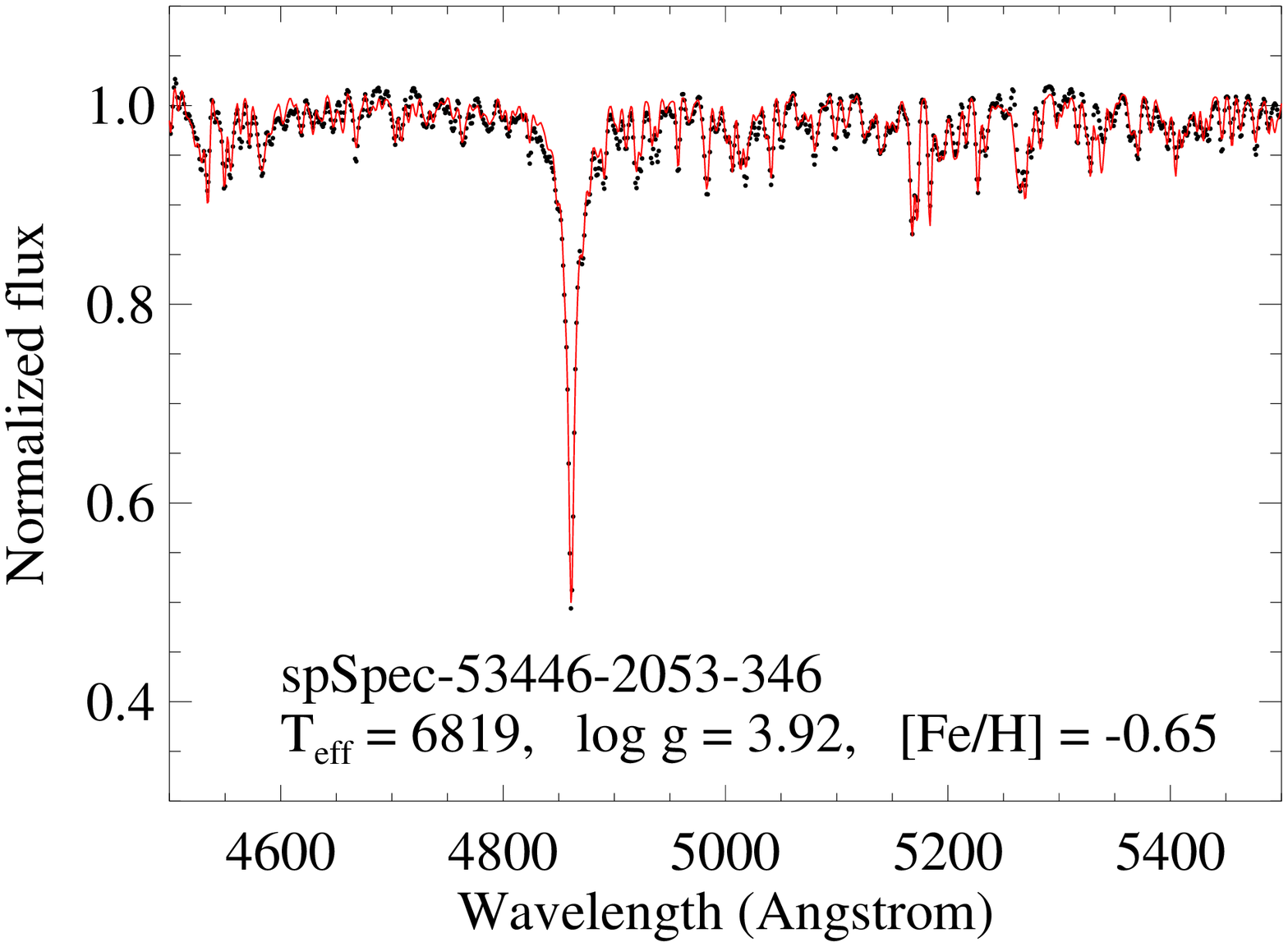}{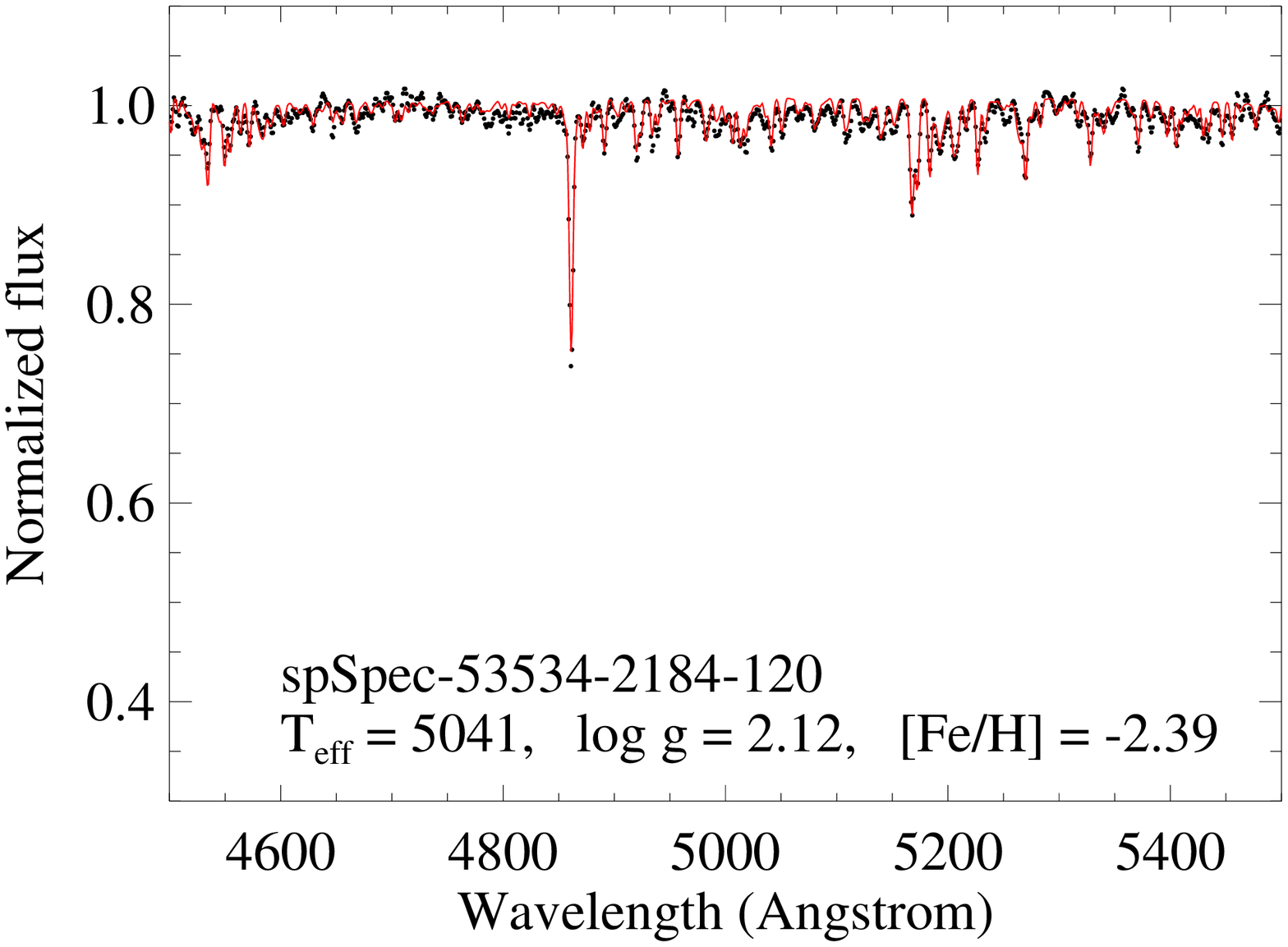}
\caption{Two examples of the results of the application of the {\tt
NGS1} grid, for a warmer, metal-rich star (left panel), and for a
cooler, metal-poor star (right panel). The black dots are the
observed data points; the red lines are synthetic spectra generated
with the atmospheric parameters adopted by the technique.}
\end{figure}
\clearpage

\begin{figure}
\centering
\includegraphics[angle=90,scale=0.65]{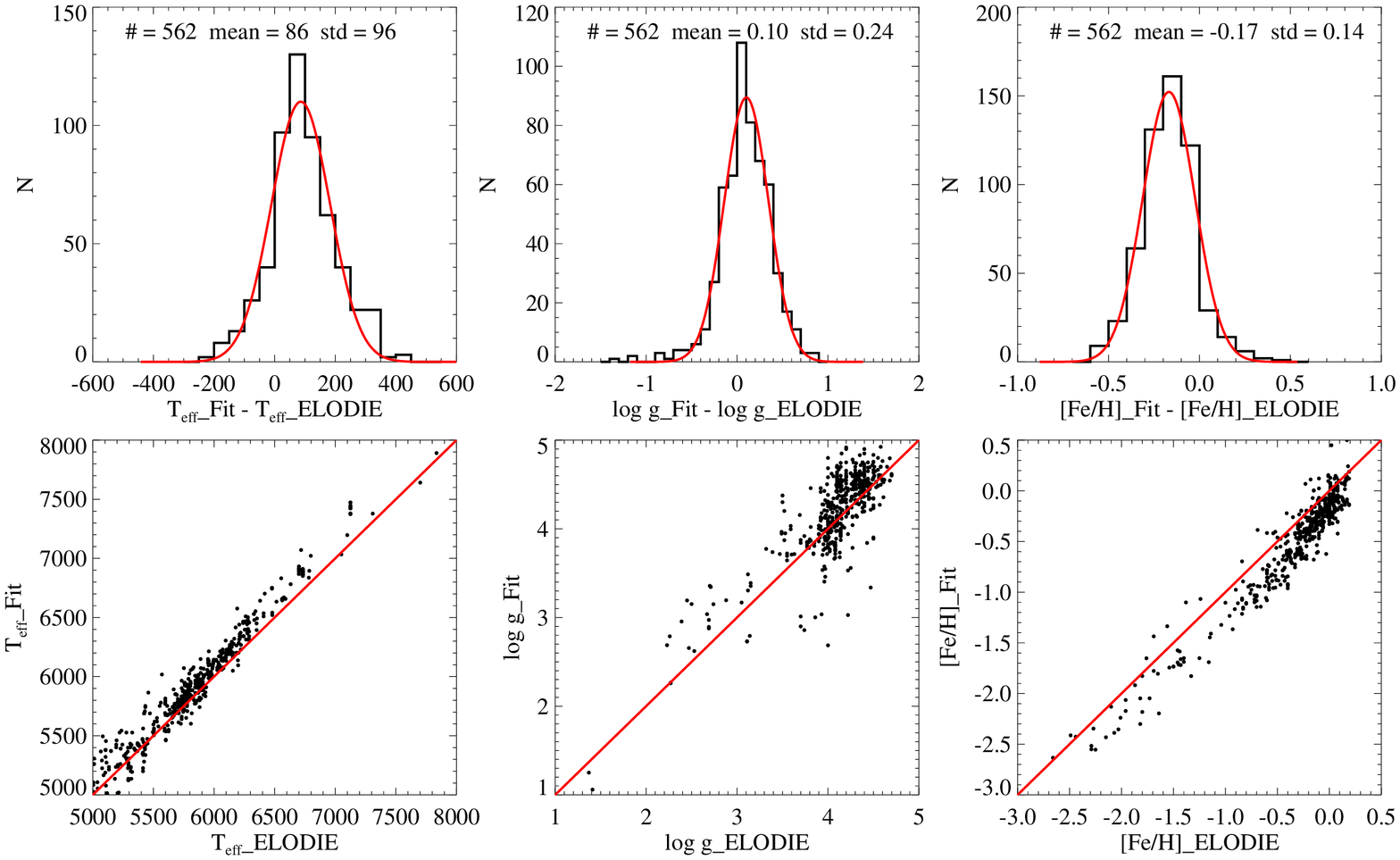}
\caption{Comparison of parameters obtained from the {\tt NGS1} grid
(FIT) with those from the ELODIE spectral library (ELODIE). It
appears that the temperature and gravity are over-estimated by 86~K
and 0.10 dex, respectively, and the metallicity is under-estimated
by 0.17 dex, based on Gaussian fits to the residuals.}
\end{figure}
\clearpage

\begin{figure}
\centering
\includegraphics[angle=90,scale=0.65]{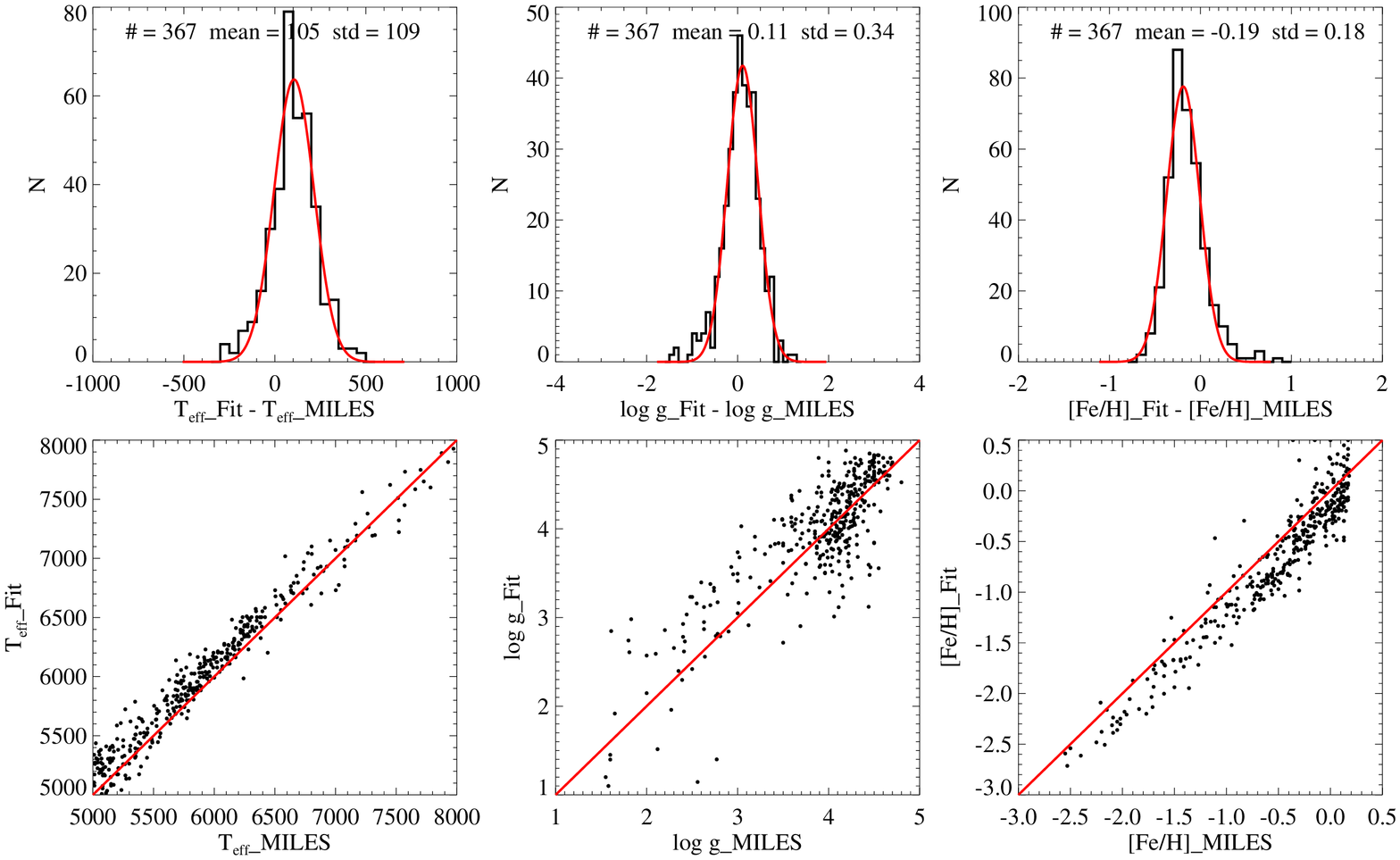}
\caption{Comparison of parameters obtained from the {\tt NGS1} grid
(FIT) with those from the MILES spectral library (MILES). It appears
that the temperature and gravity are over-estimated by 105~K and
0.11 dex, respectively, and the metallicity is under-estimated by
0.19 dex, based on Gaussian fits to the residuals.}
\end{figure}
\clearpage

\begin{figure}
\centering
\includegraphics[angle=90,scale=0.65]{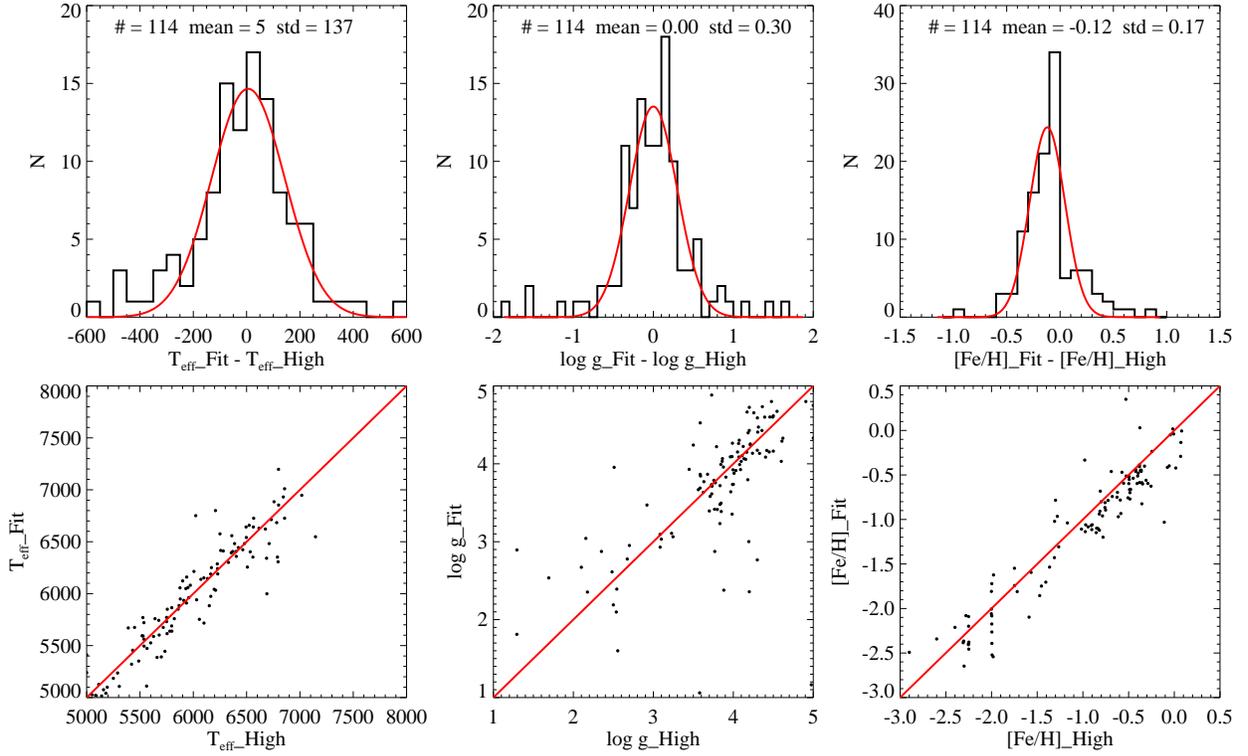}
\caption{Comparison of parameters obtained from the {\tt NGS1} grid
(FIT) with those from the analysis of high-resolution spectroscopy
of SDSS-I/SEGUE stars (High). The parameters from the
high-resolution data are averages of two independent analyses.
Unlike the results from comparisons with the ELODIE and MILES
spectral libraries, the effective temperature and the surface
gravity present almost no offsets.}
\end{figure}
\clearpage

\begin{figure}
\centering
\includegraphics[scale=0.70]{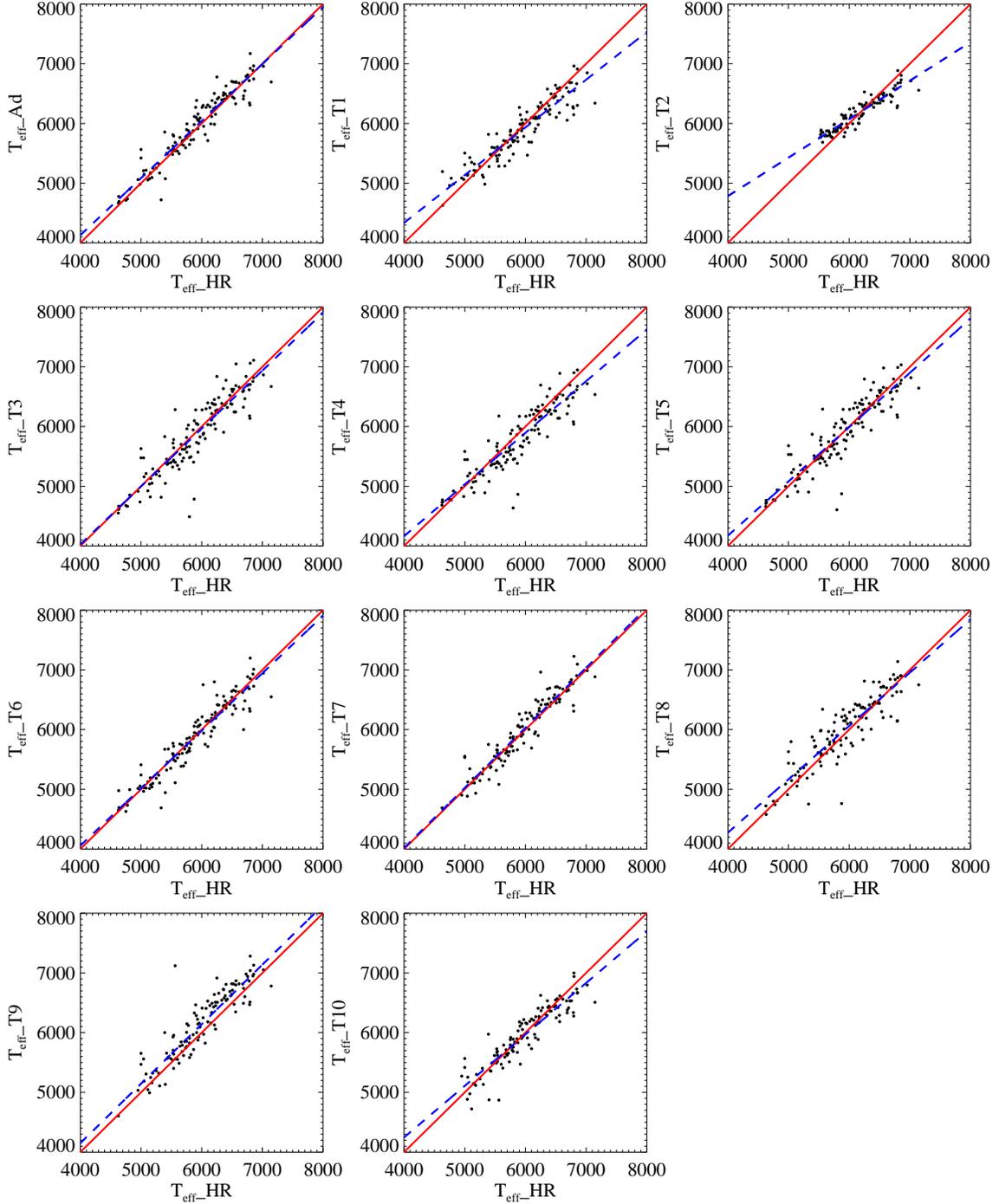}
\caption{Comparison of effective temperatures estimated from
individual methods with those from the analysis of high-resolution
spectra of SDSS-I/SEGUE stars. `HR' indicates the high-resolution
analysis results. The red solid line is the one-to-one correlation
line; the blue dashed line is the least squares fit to the data.}
\end{figure}
\clearpage

\begin{figure}
\centering
\includegraphics[scale=0.75]{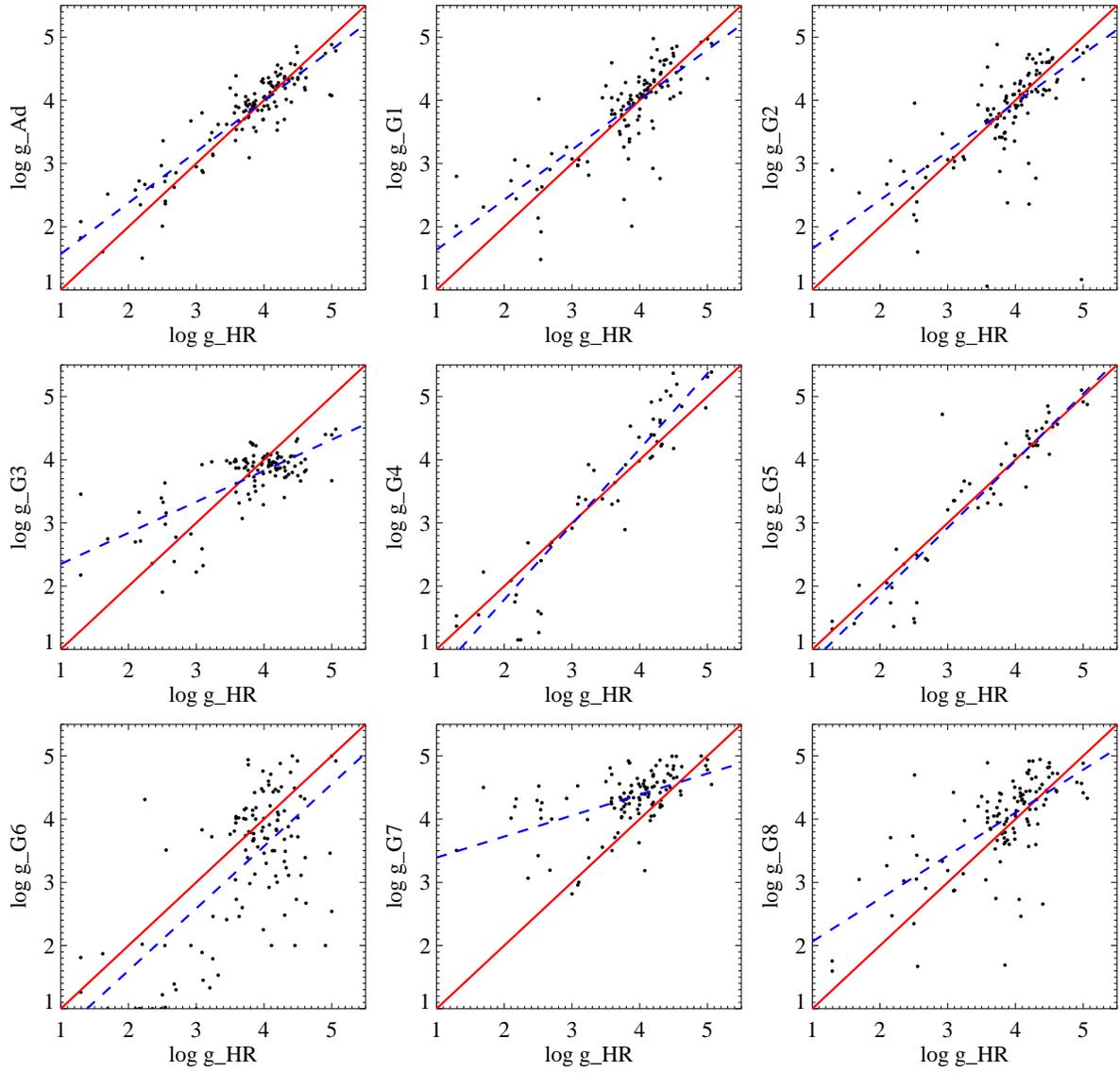}
\caption{Same as Fig. 9, but for surface gravities.}
\end{figure}
\clearpage

\begin{figure}
\centering
\includegraphics[scale=0.7]{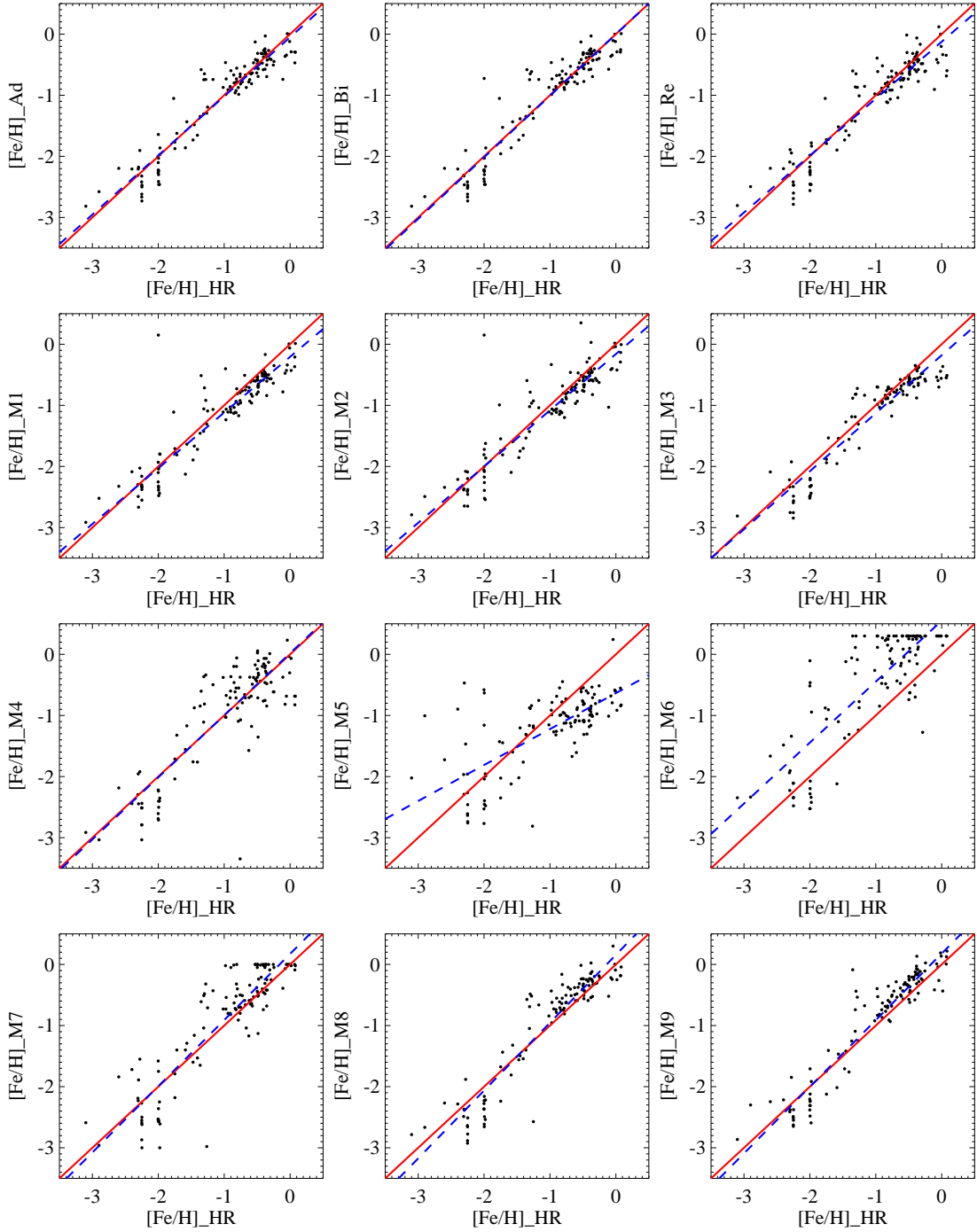}
\caption{Same as Fig. 9, but for metallicities.}
\end{figure}
\clearpage

\begin{figure}
\centering
\includegraphics[scale=0.80]{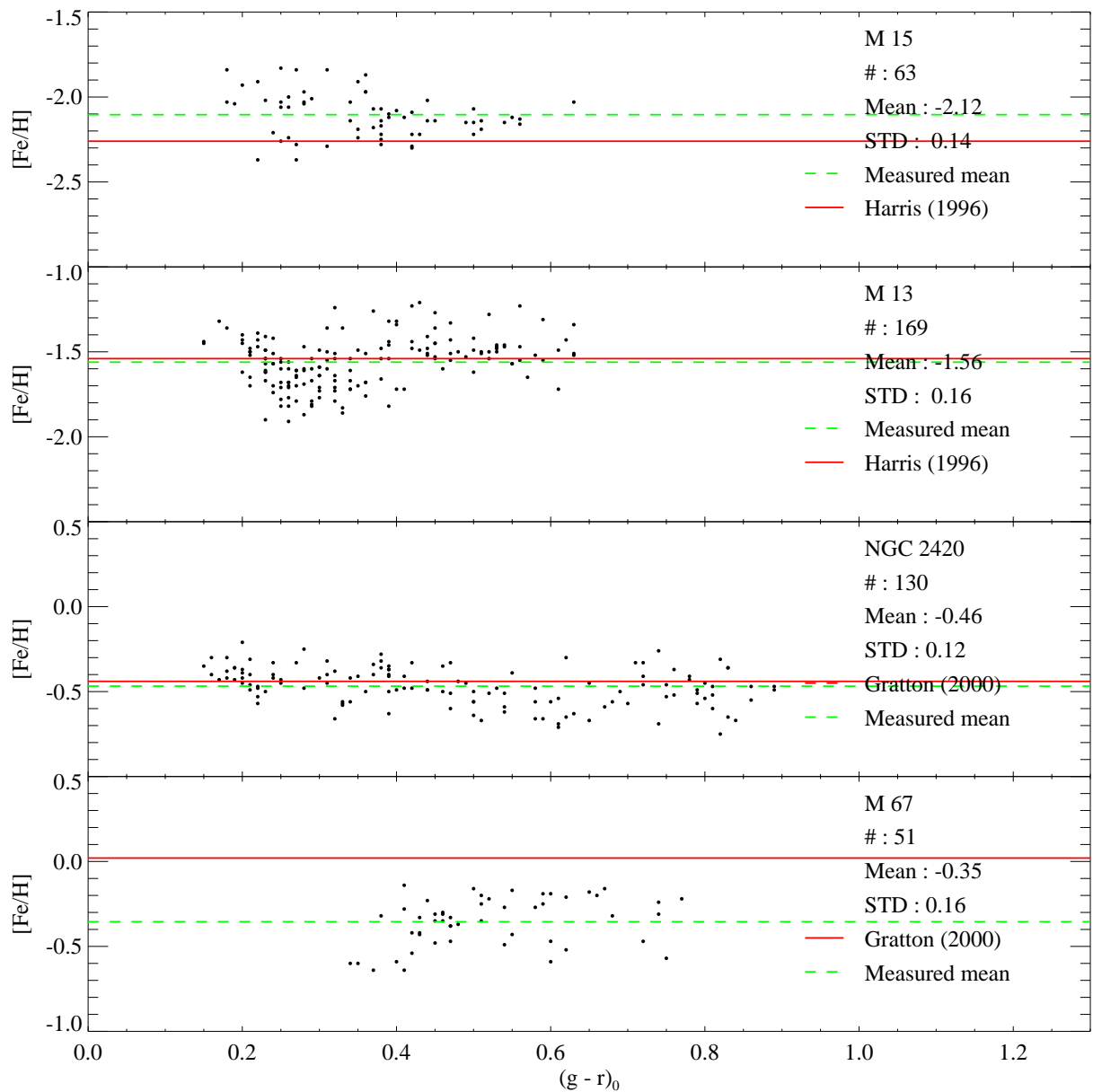}
\caption{SSPP estimated metallicities with respect to $g-r$ for
member stars of M~15, M~13, NGC~2420, and M~67. The green dashed
line is the Gaussian average of the likely member stars of each
cluster. See Paper II for a more detailed discussion of
the selection of member stars and calculation of the Gaussian means. The red
line is the adopted literature value. M~15 and M~13 are taken from Harris
(1996); NGC~2420 and M~67 are taken from Gratton (2000).}
\end{figure}
\clearpage

\begin{deluxetable}{lrccc}
\tablecolumns{5} \tablewidth{0pc}
\renewcommand{\tabcolsep}{3pt} \tablecaption{Summary of High-Resolution
Spectroscopy for SDSS and SEGUE Stars}
\tablehead{\colhead{Telescope} & \colhead{Instrument} &
\colhead{Resolving} & \colhead{Wavelength} &
\colhead{Number} \\
\colhead{} & \colhead{} & \colhead{power} & \colhead{coverage (\AA)}
& \colhead{of stars} } \startdata
Keck - I &  HIRES & 45000 &  3800$-$10000  & 11 \\
Keck - II&  ESI   & 6000  &  3800$-$10000  & 25 \\
HET      & HRS    & 15000 &  4400$-$8000  & 110 \\
Subaru   & HDS    & 45000 &  3000$-$8000  & 9 \\
\enddata
\end{deluxetable}

\clearpage

\begin{deluxetable}{rrrrrrrrr}
\tablecolumns{9} \tablewidth{0pc} \tablecaption{Line Band and Sideband Widths and Format of the Output from the
SSPP} \tablehead{\colhead{Column} & \colhead{Format} &
\colhead{Description} & \colhead{Central} & \colhead{Width} &
\colhead{Red}& \colhead{Width} & \colhead{Blue}& \colhead{Width} \\
\colhead{} & \colhead{} & \colhead{} & \colhead{(\AA)} &
\colhead{(\AA)} & \colhead{(\AA)}& \colhead{(\AA)} &
\colhead{(\AA)}& \colhead{(\AA)}} \startdata
 1   &    A22   &    spSpec name   &  \dots & \dots&  \dots  & \dots&  \dots  &\dots  \\
 2   &    F8.3  &           H8 (3) &  3889.0&   3.0&   3912.0&   8.0&   3866.0&   8.0\\
 3   &    F8.3  &          H8 (12) &  3889.1&  12.0&   4010.0&  20.0&   3862.0&  20.0\\
 4   &    F8.3  &          H8 (24) &  3889.1&  24.0&   4010.0&  20.0&   3862.0&  20.0\\
 5   &    F8.3  &          H8 (48) &  3889.1&  48.0&   4010.0&  20.0&   3862.0&  20.0\\
 6   &    F8.3  &    Ca II K12 (12)&  3933.7&  12.0&   4010.0&  20.0&   3913.0&  20.0\\
 7   &    F8.3  &    Ca II K18 (18)&  3933.7&  18.0&   4010.0&  20.0&   3913.0&  20.0\\
 8   &    F8.3  &    Ca II  K6 (6) &  3933.7&   6.0&   4010.0&  20.0&   3913.0&  20.0\\
 9   &    F8.3  &          Ca II K &  3933.6&  30.0&   4010.0&   5.0&   3910.0&   5.0\\
10   &    F8.3  &        Ca II HK+ &  3962.0&  75.0&   4010.0&   5.0&   3910.0&   5.0\\
11   &    F8.3  &      H$\epsilon$ &  3970.0&  50.0&   4010.0&   5.0&   3910.0&   5.0\\
12   &    F8.3  &    Ca II K16 (16)&  3933.7&  16.0&   4018.0&  20.0&   3913.0&  10.0\\
13   &    F8.3  &            Sr II &  4077.0&   8.0&   4090.0&   6.0&   4070.0&   4.0\\
14   &    F8.3  &             He I &  4026.2&  12.0&   4154.0&  20.0&   4010.0&  20.0\\
15   &    F8.3  &   H$\delta$ (12) &  4101.8&  12.0&   4154.0&  20.0&   4010.0&  20.0\\
16   &    F8.3  &   H$\delta$ (24) &  4101.8&  24.0&   4154.0&  20.0&   4010.0&  20.0\\
17   &    F8.3  &   H$\delta$ (48) &  4101.8&  48.0&   4154.0&  20.0&   4010.0&  20.0\\
18   &    F8.3  &        H$\delta$ &  4102.0&  64.0&   4154.0&  20.0&   4010.0&  20.0\\
19   &    F8.3  &             Ca I &  4226.0&   4.0&   4232.0&   4.0&   4211.0&   6.0\\
20   &    F8.3  &        Ca I (12) &  4226.7&  12.0&   4257.0&  20.0&   4154.0&  20.0\\
21   &    F8.3  &        Ca I (24) &  4226.7&  24.0&   4257.0&  20.0&   4154.0&  20.0\\
22   &    F8.3  &         Ca I (6) &  4226.7&   6.0&   4257.0&  20.0&   4154.0&  20.0\\
23   &    F8.3  &      CH G-band   &  4305.0&  15.0&   4367.0&  10.0&   4257.0&  20.0\\
24   &    F8.3  &   H$\gamma$ (12) &  4340.5&  12.0&   4425.0&  20.0&   4257.0&  20.0\\
25   &    F8.3  &   H$\gamma$ (24) &  4340.5&  24.0&   4425.0&  20.0&   4257.0&  20.0\\
26   &    F8.3  &   H$\gamma$ (48) &  4340.5&  48.0&   4425.0&  20.0&   4257.0&  20.0\\
27   &    F8.3  &        H$\gamma$ &  4340.5&  54.0&   4425.0&  20.0&   4257.0&  20.0\\
28   &    F8.3  &             He I &  4471.7&  12.0&   4500.0&  20.0&   4425.0&  20.0\\
29   &    F8.3  &           G-blue &  4305.0&  26.0&   4507.0&  14.0&   4090.0&  12.0\\
30   &    F8.3  &          G-whole &  4321.0&  28.0&   4507.0&  14.0&   4096.0&  12.0\\
31   &    F8.3  &            Ba~II &  4554.0&   6.0&   4560.0&   4.0&   4538.0&   4.0\\
32   &    F8.3  & C$_{12}$C$_{13}$ &  4737.0&  36.0&   4770.0&  20.0&   4423.0&  10.0\\
33   &    F8.3  &             CC12 &  4618.0& 256.0&   4780.0&   5.0&   4460.0&  10.0\\
34   &    F8.3  &          Metal-1 &  4584.0& 442.0&   4805.8&   5.0&   4363.0&   5.0\\
35   &    F8.3  &    H$\beta$ (12) &  4862.3&  12.0&   4905.0&  20.0&   4790.0&  20.0\\
36   &    F8.3  &    H$\beta$ (24) &  4862.3&  24.0&   4905.0&  20.0&   4790.0&  20.0\\
37   &    F8.3  &    H$\beta$ (48) &  4862.3&  48.0&   4905.0&  20.0&   4790.0&  20.0\\
38   &    F8.3  &         H$\beta$ &  4862.3&  60.0&   4905.0&  20.0&   4790.0&  20.0\\
39   &    F8.3  &          C$_{2}$ &  5052.0& 204.0&   5230.0&  20.0&   4935.0&  10.0\\
40   &    F8.3  &     C$_{2}$+Mg I &  5069.0& 238.0&   5230.0&  20.0&   4935.0&  10.0\\
41   &    F8.3  & MgH+Mg I+C$_{2}$ &  5085.0& 270.0&   5230.0&  20.0&   4935.0&  10.0\\
42   &    F8.3  &         MgH+Mg I &  5198.0&  44.0&   5230.0&  20.0&   4935.0&  10.0\\
43   &    F8.3  &              MgH &  5210.0&  20.0&   5230.0&  20.0&   4935.0&  10.0\\
44   &    F8.3  &             Cr I &  5206.0&  12.0&   5239.0&   8.0&   5197.5&   5.0\\
45   &    F8.3  &       Mg I+Fe II &  5175.0&  20.0&   5240.0&  10.0&   4915.0&  10.0\\
46   &    F8.3  &             Mg I &  5183.0&   2.0&   5240.0&  10.0&   4915.0&  10.0\\
47   &    F8.3  &             Mg I &  5170.5&  12.0&   5285.0&  20.0&   5110.0&  20.0\\
48   &    F8.3  &             Mg I &  5176.5&  24.0&   5285.0&  20.0&   5110.0&  20.0\\
49   &    F8.3  &             Mg I &  5183.5&  12.0&   5285.0&  20.0&   5110.0&  20.0\\
50   &    F8.3  &             Na I &  5890.0&  20.0&   5918.0&   6.0&   5865.0&  10.0\\
51   &    F8.3  &          Na (12) &  5892.9&  12.0&   5970.0&  20.0&   5852.0&  20.0\\
52   &    F8.3  &          Na (24) &  5892.9&  24.0&   5970.0&  20.0&   5852.0&  20.0\\
53   &    F8.3  &   H$\alpha$ (12) &  6562.8&  12.0&   6725.0&  50.0&   6425.0&  50.0\\
54   &    F8.3  &   H$\alpha$ (24) &  6562.8&  24.0&   6725.0&  50.0&   6425.0&  50.0\\
55   &    F8.3  &   H$\alpha$ (48) &  6562.8&  48.0&   6725.0&  50.0&   6425.0&  50.0\\
56   &    F8.3  &   H$\alpha$ (70) &  6562.8&  70.0&   6725.0&  50.0&   6425.0&  50.0\\
57   &    F8.3  &              CaH &  6788.0& 505.0&   7434.0&  10.0&   6532.0&   5.0\\
58   &    F8.3  &              TiO &  7209.0& 333.3&   7434.0&  10.0&   6532.0&   5.0\\
59   &    F8.3  &               CN &  6890.0&  26.0&   7795.0&  10.0&   6870.0&  10.0\\
60   &    F8.3  &         O~I tri  &  7775.0&  30.0&   7805.0&  10.0&   7728.0&  10.0\\
61   &    F8.3  &              K I &  7687.0&  34.0&   8080.0&  10.0&   7510.0&  10.0\\
62   &    F8.3  &              K I &  7688.0&  95.0&   8132.0&   5.0&   7492.0&   5.0\\
63   &    F8.3  &             Na I &  8187.5&  15.0&   8190.0&  55.0&   8150.0&  10.0\\
64   &    F8.3  &         Na~I-red &  8190.2&  33.0&   8248.6&   5.0&   8140.0&   5.0\\
65   &    F8.3  &       Ca II (26) &  8498.0&  26.0&   8520.0&  10.0&   8467.5&  25.0\\
66   &    F8.3  &     Paschen (13) &  8467.5&  13.0&   8570.0&  14.0&   8457.0&  10.0\\
67   &    F8.3  &            Ca II &  8498.5&  29.0&   8570.0&  14.0&   8479.0&  10.0\\
68   &    F8.3  &       Ca II (40) &  8542.0&  40.0&   8570.0&  14.0&   8479.0&  10.0\\
69   &    F8.3  &            Ca II &  8542.0&  16.0&   8600.0&  60.0&   8520.0&  20.0\\
70   &    F8.3  &     Paschen (42) &  8598.0&  42.0&   8630.5&  23.0&   8570.0&  14.0\\
71   &    F8.3  &            Ca II &  8662.1&  16.0&   8694.0&  12.0&   8600.0&  60.0\\
72   &    F8.3  &       Ca II (40) &  8662.0&  40.0&   8712.5&  25.0&   8630.5&  23.0\\
73   &    F8.3  &     Paschen (42) &  8751.0&  42.0&   8784.0&  16.0&   8712.5&  25.0\\
74   &    F7.3  &             TiO1 &  6720.5&   5.0&   6720.5&   5.0&   6705.5&   5.0\\
75   &    F7.3  &             TiO2 &  7059.5&   5.0&   7059.5&   5.0&   7044.5&   5.0\\
76   &    F7.3  &             TiO3 &  7094.5&   5.0&   7094.5&   5.0&   7081.5&   5.0\\
77   &    F7.3  &             TiO4 &  7132.5&   5.0&   7132.5&   5.0&   7117.5&   5.0\\
78   &    F7.3  &             TiO5 &  7130.5&   9.0&   7130.5&   9.0&   7044.0&   4.0\\
79   &    F7.3  &             CaH1 &  6385.0&  10.0&   6415.0&  10.0&   6350.0&  10.0\\
80   &    F7.3  &             CaH2 &  6830.0&  32.0&   6830.0&  32.0&   7044.0&   4.0\\
81   &    F7.3  &             CaH3 &  6975.0&  30.0&   6975.0&  30.0&   7044.0&   4.0\\
82   &    F7.3  &             CaOH &  6235.0&  10.0&   6235.0&  10.0&   6349.5&   9.0\\
83   &    F7.3  &        H$\alpha$ &  6563.0&   6.0&   6563.0&   6.0&   6550.0&  10.0\\
84   &    F6.1  &          $<S/N>$ &        &      &         &      &         &      \\
85   &    A10   &          RV Flag &        &      &         &      &         &      \\
\enddata
\tablecomments{$<S/N>$ is the average signal to noise ratio per
pixel over 3850\,{\AA} to 6000\,{\AA}.}
\end{deluxetable}

\clearpage

\begin{deluxetable}{clrccccccc}
\tablecolumns{10} \tablewidth{0pc}
\renewcommand{\tabcolsep}{2pt}
\tablecaption{Comparison of Parameters from {\tt NGS1} and {\tt
NGS2} grids with the ELODIE and MILES Libraries and 
High-Resolution Values} \tablehead{\colhead{} & \colhead{} &
\colhead{} & \colhead{} &
\multicolumn{2}{c}{$T_{\rm eff}$} & \multicolumn{2}{c}{$\log g$} & \multicolumn{2}{c}{[Fe/H]} \\
\colhead{Grid} & \colhead{Library} & \colhead{$S/N$} & \colhead{N} &
\colhead{$<\Delta>$} & \colhead{$\sigma$} & \colhead{$<\Delta>$} &
\colhead{$\sigma$} & \colhead{$<\Delta>$} & \colhead{$\sigma$} \\
\colhead{} & \colhead{} & \colhead{} & \colhead{} & \colhead{(K)} &
\colhead{(K)} & \colhead{(dex)} & \colhead{(dex)} & \colhead{(dex)}&
\colhead{(dex)}} \startdata
{\tt NGS1} &         &       &     &       &     &         &      &         &      \\
           & ELODIE  & Full  & 562 &  86   &  96 & $+0.10$ & 0.24 & $-0.17$ & 0.14 \\
           & ELODIE  & 50/1  & 543 & 110   & 104 & $-0.01$ & 0.28 & $-0.20$ & 0.13 \\
           & ELODIE  & 25/1  & 538 & 114   & 118 & $-0.05$ & 0.32 & $-0.18$ & 0.16 \\
           & ELODIE  &12.5/1 & 489 & 146   & 155 & $-0.11$ & 0.43 & $+0.20$ & 0.20 \\
           & ELODIE  &6.25/1 & 370 & 149   & 225 & $-0.16$ & 0.67 & $+0.47$ & 0.19 \\
           & MILES   & Full  & 367 & 105   & 109 & $+0.11$ & 0.34 & $-0.19$ & 0.18 \\
           & HIGHRES & Full  & 114 & $+5$  & 137 & $+0.00$ & 0.30 & $-0.12$ & 0.17 \\[+5pt]
\hline \\[-5pt]
{\tt NGS2} &         &       &     &      &     &         &      &         &      \\
           & ELODIE  & Full  & 557 &\dots &\dots& $+0.12$ & 0.26 & $-0.23$ & 0.16 \\
           & MILES   & Full  & 341 &\dots &\dots& $+0.14$ & 0.30 & $-0.25$ & 0.15 \\
           & HIGHRES & Full  & 112 &\dots &\dots& $+0.09$ & 0.27 & $-0.14$ & 0.17 \\
\enddata
\tablecomments{HIGHRES is the average of the two high-resolution
analyses.}
\end{deluxetable}

\clearpage

\begin{deluxetable}{cl}
\tablecolumns{2} \tablewidth{0pc} \tablecaption{Brief Descriptions
of SSPP Flags} \tablehead{\colhead{Flag} & \colhead{Comment}}
\startdata
n & All appears normal \\
D & Likely white dwarf \\
d & Likely sdO or sdB \\
H & Hot star with $T_{\rm eff} >$ 10000 K\\
h & Helium line detected, possibly very hot star\\
l & Likely solar abundance, late-type star\\
E & Emission lines in spectrum \\
S & Sky spectrum \\
V & No radial velocity information available\\
N & Noisy spectrum at extrema \\
C & The photometric $g-r$ color may be incorrect \\
B & Unexpected H$\alpha$ line strength predicted from H$\delta$ \\
G & Strong G-band feature \\
g & Mild G-band feature \\
\enddata
\end{deluxetable}

\clearpage

\begin{deluxetable}{ccccccccccc}
\tablecolumns{10}  \tablewidth{0in} \tabletypesize{\scriptsize}
\renewcommand{\tabcolsep}{3pt} \tablecaption{Valid Ranges of Effective
Temperature, $g-r$ Color, and $S/N$ for Individual Methods in the
SSPP} \tablehead{\colhead{Temperature} & \colhead{} & \colhead{} &
\colhead{} & \colhead{} & \colhead{} & \colhead{} & \colhead{} &
\colhead{} & \colhead{}} \startdata
Method    & T1 & T2 & T3 & T4 & T5 &   T6 &        T7 & T8         &      T9 &     T10 \\
          & $HA24$ & $HD24$ & K-MOD & G-ISO & EMP   &      {\tt NGS1} &        ANN & WBG99       &    {\tt k24} &  {\tt ki13} \\
$T_{\rm eff}$\tablenotemark{1} & 4.5 $-$ 8.5  & 5.5 $-$ 8.5 &4 $-$ 10 &4 $-$ 10 & 4 $-$ 8&4 $-$ 8 & 5 $-$ 7.5 & 4.5 $-$ 10 & 5 $-$ 7 & 5 $-$ 7 \\
$g-r$    &   0.0 $-$ 0.8 & 0.0 $-$ 0.6 &-0.2 $-$ 1.2 &0.0 $-$ 1.2 &
-0.2 $-$ 1.2 & 0.0 $-$ 1.2 & 0.1 $-$ 0.7 & $-$0.2 $-$ 0.8 & 0.1 $-$ 0.7 & 0.1 $-$ 0.7  \\
$S/N$ & \dots &  \dots & \dots & \dots & \dots  & $>$ 10.0 & $>$ 10.0 & $>$ 10.0 & $>$ 10.0 & $>$ 10.0 \\
$S/N$ ($g-r < 0.3$) & \dots & \dots & \dots &  \dots & \dots & $>$ 20.0\tablenotemark{2} & $>$ 15.0 & $>$ 15.0 & $>$ 15.0 & $>$ 15.0 \\[+5pt]
\hline \\[-5pt]
Gravity & & & & & & & & & \\[+5pt]
\hline \\
Method & G1 & G2 & G3 & G4 & G5 & G6 & G7 & G8 &  &  \\
       & {\tt NGS2} & {\tt NGS1}  & ANN & CaI & MgH & WBG & {\tt k24} &  {\tt ki13} & & \\
$T_{\rm eff}$\tablenotemark{1} & 5 $-$ 8 & 5 $-$ 8 & 5 $-$ 7.5 & 4.5
$-$ 6 & 4.5 $-$ 6 & 4.5 $-$ 10 & 5 $-$ 7 & 5 $-$ 7 & & \\
$g-r$ & 0.0 $-$ 0.7 & 0.0 $-$ 0.7 & 0.1 $-$ 0.7 & 0.4 $-$ 0.9 &
0.4 $-$ 0.9 & $-$0.2 $-$ 0.8 & 0.1 $-$ 0.7 & 0.1 $-$ 0.7 &  & \\
$S/N$ & $>$ 10.0 & $>$ 10.0 & $>$ 10.0 & $>$ 10.0 & $>$ 10.0 & $>$ 10.0 & $>$10.0 & $>$ 10.0 & & \\
$S/N$ ($g-r <$ 0.3)& $>$ 20.0\tablenotemark{2} & $>$ 20.0\tablenotemark{2} & $>$ 15.0 & \dots & \dots & $>$ 15.0 & $>$ 15.0& $>$ 15.0 & & \\[+5pt]
\hline \\[-5pt]
Metallicity & & & & & & & & & \\[+5pt]
\hline \\[-5pt]
Method & M1 & M2 & M3 & M4 & M5 & M6 & M7& M8 & M9 & \\
       & {\tt NGS2} & {\tt NGS1} & ANN & CaII K & ACF & CaII T & WBG& {\tt k24} &  {\tt ki13} & \\
$T_{\rm eff}$\tablenotemark{1} & 4 $-$ 8 & 4 $-$ 8 & 5 $-$ 7.5 & 5
$-$ 7
& \dots & \dots & 4.5 $-$ 10 & 5 $-$ 7 & 5 $-$ 7 & \\
$g-r$ & 0.0 $-$ 1.2 & 0.0 $-$ 1.2 & 0.1 $-$ 0.7 & 0.1 $-$ 0.7 &
\dots &
\dots& $-$0.2 $-$ 0.8 & 0.1 $-$ 0.7 & 0.1 $-$ 0.7 &\\
$S/N$ & $>$ 10.0 & $>$ 10.0 & $>$ 10.0 & $>$ 10.0 & \dots & \dots & $>$ 10.0 & $>$ 10.0 & $>$ 10.0 &\\
$S/N$ ($g-r <$ 0.3) & $>$ 20.0 \tablenotemark{2}& $>$ 20.0 \tablenotemark{2}& $>$ 15.0 & $>$ 15.0 & \dots & \dots & $>$ 15.0 & $>$ 15.0 & $>$ 15.0 &\\
\enddata
\tablenotetext{1}{$T_{\rm eff}$ is in units of 1000 K.}
\tablenotetext{2}{In hhis case, $g-r < 0.4$ } \tablecomments{$S/N$
is the average signal to noise ratio per pixel over 3850\,{\AA} to
6000\,{\AA}. $HA24$ and $HD24$ are the temperature estimates from
the H$\alpha$ and H$\delta$ line index in 24\,{\AA} widths,
respectively. The temperature estimated from Kurucz models is
referred to as K-MOD; G-ISO is for the Girardi et al. (2004)
isochrones. EMP is the temperature determined by equation 12. These
temperature estimates are used only if the color flag is not raised
and the mean of these five temperature estimates are greater than
7500~K or less than 4500~K. ANN is the neural network approach. ACF
is the Autocorrelation Function method, and Ca~II~T is based on the
Ca II triplet line index. [Fe/H] estimates from these approaches are
not used at by the SSPP at present. WBG99 is the method from
Wilhelm, Beers, \& Gray (1999).}
\end{deluxetable}

\clearpage

\begin{deluxetable}{rcrrrrrrrrrrr}
\tablecolumns{13} \tablewidth{0in}
\renewcommand{\tabcolsep}{2pt} \tablecaption{Comparison of $T_{\rm eff}$
Estimates from Individual Methods with Those from Two
High-Resolution Analyses} \tablehead{ \colhead{} & \colhead{} &
\colhead{AD} & \colhead{T1} & \colhead{T2} & \colhead{T3} &
\colhead{T4} & \colhead{T5} & \colhead{T6} & \colhead{T7} &
\colhead{T8} & \colhead{T9} & \colhead{T10}} \startdata

HA1 & N &  81  &    81  &    81  &    81  &    81  &    81  &    81  &    74  &    81  &    76  &    76  \\
    & $<\Delta>$ & $+$183  &    $+$93  &   $+$194  &   $+$124  &    $+$41  &   $+$142  &   $+$145  &   $+$162  &   $+$228  &   $+$329  &   $+$127  \\
    & $\sigma$   & 169  &   138  &   186  &   285  &   244  &   259  &   171  &   186  &   194  &   223  &   202  \\[+5pt]
\hline\\[-5pt]
HA2 & N & 124  &   123  &   125  &   125  &   125  &   125  &   124  &   111  &   122  &   115  &   115  \\
    & $<\Delta>$ & $-$30  &  $-$137  &   $-$27  &  $-$108  &  $-$180  &   $-$82  &   $-$74  &   $-$73  &    $-$8  &    $+$48  &  $-$100  \\
    & $\sigma$   & 145  &   191  &   312  &   163  &   207  &   186  &   177  &   149  &   200  &   177  &   182  \\[+5pt]
\hline\\[-5pt]
MEAN & N & 124  &   123  &   125  &   125  &   125  &   125  &   124  &   111  &   122  &   115  &   115  \\
    & $<\Delta>$ &  $+$55  &   $-$18  &    $+$74  &   $-$41  &   $-$85  &    $-$4  &    $+$10  &    $+$28  &    $+$85  &   $+$148  &    $-$6  \\
    & $\sigma$   & 123  &   158  &   228  &   198  &   206  &   188  &   133  &   144  &   182  &   203  &   167  \\

\enddata
\tablecomments{`AD' is the adopted estimate of $T_{\rm eff}$. `HA1'
indicates the analysis performed by C.A.; `HA2' indicates the
analysis performed by T.S.. `MEAN' is the average of the two
analyses. `N' is the number of stars compared. $<\Delta>$ is the
mean from a Gaussian fit to the residuals of $T_{\rm eff}$ between
the SSPP and the high-resolution analysis; $\sigma$ is the standard
deviation of the fit.}
\end{deluxetable}

\clearpage

\begin{deluxetable}{rcrrrrrrrrr}
\tablecolumns{11} \tablewidth{0in}
\renewcommand{\tabcolsep}{2pt}
\tablecaption{Comparison of log $g$ Estimates from Individual
Methods with Those from Two High-Resolution Analyses}
\tablehead{\colhead{} & \colhead{} & \colhead{AD} & \colhead{G1} &
\colhead{G2} & \colhead{G3} & \colhead{G4} & \colhead{G5} &
\colhead{G6} & \colhead{G7} & \colhead{G8}} \startdata

HA1 & N &  81    &    74  &    76  &    74  &    34  &    34  &    81  &    76  &    76  \\
    & $<\Delta>$ &  $+$0.09  &    $+$0.15  &    $+$0.07  &   $-$0.16  &    $+$0.21  &   $-$0.02  &   $-$0.31  &    $+$0.37  &    $+$0.25  \\
    & $\sigma$   &  0.23  &    0.19  &    0.18  &    0.29  &    0.40  &    0.13  &    0.93  &    0.33  &    0.30  \\[+5pt]
\hline\\[-5pt]
HA2 & N &  124  &   114  &   116  &   111  &    54  &    54  &   122  &   115  &   115  \\
    & $<\Delta>$ &  $+$0.03  &    $+$0.04  &   $-$0.01  &   $-$0.18  &    $+$0.13  &    $+$0.02  &   $-$0.26  &    $+$0.31  &    $+$0.13  \\
    & $\sigma$   &  0.32  &    0.37  &    0.38  &    0.55  &    0.42  &    0.28  &    0.87  &    0.37  &    0.44  \\[+5pt]
\hline\\[-5pt]

MEAN & N &  124  &   114  &   116  &   111  &    54  &    54  &   122  &   115  &   115  \\
    & $<\Delta>$ &  $+$0.04  &    $+$0.08  &    $+$0.01  &   $-$0.16  &    $+$0.15  &    $+$0.04  &   $-$0.29  &    $+$0.35  &    $+$0.18  \\
    & $\sigma$   &  0.25  &    0.26  &    0.29  &    0.42  &    0.40  &    0.27  &    0.89  &    0.33  &    0.41  \\

\enddata
\end{deluxetable}

\clearpage

\begin{deluxetable}{rcrrrrrrrrrrrr}
\tablecolumns{14} \tablewidth{0in}
\renewcommand{\tabcolsep}{2pt}
\tablecaption{Comparison of [Fe/H] Estimates from Individual Methods
with Those from Two High-Resolution Analyses}
\tablehead{\colhead{}
& \colhead{} & \colhead{AD} & \colhead{BI} & \colhead{RE} &
\colhead{M1} & \colhead{M2} & \colhead{M3} & \colhead{M4} &
\colhead{M5} & \colhead{M6} & \colhead{M7} & \colhead{M8} &
\colhead{M9}} \startdata

HA1 & N &  81  &    81  &    81  &    80  &    80  &    74  &    74  &    79  &    79  &    81  &    76  &    76  \\
    & $<\Delta>$ & $-$0.08  &   $-$0.06  &   $-$0.14  &   $-$0.23  &   $-$0.19  &   $-$0.16  &    $+$0.07  &   $-$0.47  &    $+$0.50  &    $+$0.10  &    $+$0.05  &    $+$0.12  \\
    & $\sigma$   &  0.16  &    0.16  &    0.17  &    0.13  &    0.15  &    0.18  &    0.35  &    0.31  &    0.35  &    0.24  &    0.17  &    0.15  \\[+5pt]
\hline\\[-5pt]
HA2 & N & 124  &   124  &   124  &   123  &   123  &   111  &   110  &   122  &   118  &   122  &   115  &   115  \\
    & $<\Delta>$ & $-$0.00  &    $+$0.01  &   $-$0.04  &   $-$0.07  &   $-$0.06  &   $-$0.10  &    $+$0.07  &   $-$0.30  &    $+$0.63  &    $+$0.15  &    $+$0.13  &    $+$0.15  \\
    & $\sigma$   &   0.28  &    0.27  &    0.31  &    0.23  &    0.25  &    0.30  &    0.47  &    0.39  &    0.51  &    0.43  &    0.33  &    0.30  \\[+5pt]
\hline\\[-5pt]
MEAN & N & 124  &   124  &   124  &   123  &   123  &   111  &   110  &   122  &   118  &   122  &   115  &   115  \\
    & $<\Delta>$ & $-$0.04  &   $-$0.03  &   $-$0.08  &   $-$0.13  &   $-$0.11  &   $-$0.12  &    $+$0.04  &   $-$0.35  &    $+$0.58  &    $+$0.12  &    $+$0.10  &    $+$0.14  \\
    & $\sigma$   &   0.21  &    0.21  &    0.22  &    0.17  &    0.18  &    0.23  &    0.42  &    0.39  &    0.49  &    0.34  &    0.24  &    0.20  \\

\enddata

\end{deluxetable}

\clearpage

\end{document}